\pdfoutput=1
\documentclass[PAPER, atlasdraft=false, USenglish, texlive=2025, cernpreprint, texmf, orcidlogo]{atlasdoc}
\usepackage{atlaspackage}
\usepackage{wasysym}
\usepackage{subfig}
\usepackage{tikz}
\usetikzlibrary{positioning}
\usepackage{atlasbiblatex}

\usepackage{atlasphysics}
\graphicspath{{logos/}{figures/}}

\usepackage{ANA-EXOT-2018-39-PAPER-defs}

\AtlasTitle{Search for electroweak-scale dijet resonances using trigger-level analysis with the ATLAS detector in \qty{132}{\per\femto\barn} of \protonproton collisions at $\sqrt{s}=\qty{13}{\TeV}$}

\AtlasAbstract{This \publicationType reports on a search for dijet resonances using \qty{132}{\per\femto\barn} of \protonproton collision data recorded at $\sqrt{s} = \qty{13}{\TeV}$ by the ATLAS detector at the Large Hadron Collider.
The search is performed solely on jets reconstructed within the ATLAS trigger to overcome bandwidth limitations imposed on conventional single-jet triggers, which would otherwise reject data from decays of sub-\unit{\TeV} dijet resonances.
Collision events with two jets satisfying transverse momentum thresholds of $\pT \ge \qty{85}{\GeV}$ and jet rapidity separation of $|\yStar|<0.6$ are analysed for dijet resonances with invariant masses from \qtyrange{375}{1800}{\GeV}.
A data-driven background estimate is used to model the dijet mass distribution from multijet processes.
No significant excess above the expected background is observed.
Upper limits are set at 95\% confidence level on coupling values for a benchmark leptophobic axial-vector $\Zprime$ model and on the production cross-section for a new resonance contributing a Gaussian-distributed line-shape to the dijet mass distribution.
}

\AtlasRefCode{EXOT-2018-39}

\PreprintIdNumber{CERN-EP-2025-194}

\AtlasJournal{Phys. Rev. D}
\AtlasJournalRef{\PRD 112 (2025) 092015}
\AtlasDOI{10.1103/15p2-bkg8}

\hypersetup{pdftitle={ATLAS document},pdfauthor={The ATLAS Collaboration}}

\begin{document}

\maketitle

\tableofcontents

%

% The next lines are included from the .//01intro.tex input file
%
%
\clearpage%
\section{Introduction}
\label{sec:intro}

A key strength of collider experiments is their ability to discover new particles and observe their properties.
This capability has led to many of the major discoveries of particle physics in the last four decades, notably that of the \Wboson and \Zboson bosons~\cite{UA1:1983crd,UA2:1983tsx,UA1:1983mne,UA2:1983mlz}, top quark~\cite{PhysRevLett.74.2626,PhysRevLett.74.2632}, and the Higgs boson~\cite{HIGG-2012-27,CMS-HIG-12-028} in the mass range \qtyrange{81}{173}{\GeV}, around what is known as the ``electroweak scale.''
Nevertheless, the electroweak scale remains to be fully explored.

The ATLAS experiment~\cite{PERF-2007-01} at the Large Hadron Collider (LHC)~\cite{Evans:2008zzb} is one of two experiments where the Higgs boson was discovered with approximately \qty{11}{\per\femto\barn} of \protonproton collision data.
ATLAS recorded a factor of $\sim 13$ times more data during Run~2 (2015--2018), providing an unprecedented data sample to search for evidence of additional particles with masses at and above the electroweak scale.
Dijet resonances constitute a primary search signature for new particles.
For example, any mediator particle capable of being produced directly in \protonproton collisions should be able to decay into a dijet signature.
Searches for dijet resonances, however, have been impeded by the large background from multijet processes in the Standard Model (SM), which are produced with rates that overwhelm the normal data-taking capabilities of the ATLAS and CMS~\cite{CMS-CMS-00-001} detectors.
Consequently, it has been difficult for LHC searches to constrain electroweak-scale dijet resonances with the same coupling sensitivity as at multi-\unit{\TeV} masses.

Both ATLAS and CMS have targeted sub-\unit{\TeV} masses with new dijet searches using their large samples of Run-2 collisions.
In 2018, ATLAS published a search for dijet resonances using an innovative technique, trigger-level analysis (TLA), on one quarter of that sample~\cite{EXOT-2016-20}.
The TLA technique overcomes the normal data-taking limitations by recording the data as reconstructed and reduced by the ATLAS trigger system, rather than the full detector data stream, for subsequent analysis.
With much greater statistical power than available in traditional data-taking, the TLA technique allowed ATLAS to constrain dijet resonances at masses from \qtyrange{450}{1800}{\GeV}, far below the minimum mass of \qty{1.5}{\TeV} probed by the conventional ATLAS dijet searches in Run~2~\cite{EXOT-2019-03}.
A similar technique to TLA (``data scouting'') was prototyped by CMS with data at \qty{7}{\TeV} centre-of-mass energy and employed in CMS dijet analyses, constraining resonances in the mass range \qtyrange{500}{1800}{\GeV}~\cite{CMS-EXO-16-056,CMS-EXO-14-005,CMS-EXO-19-004}.
Similarly, LHCb~\cite{LHCb:2008vvz} employs the ``Turbo stream'' as one of their main data-taking strategies~\cite{Aaij:2016rxn,Aaij:2019uij}.
Searches for dijet resonances in complementary channels were performed with additional requirements on the dijet system, such as large transverse momentum (\pT)~\cite{EXOT-2017-01,EXOT-2022-33}, photon or jet initial-state radiation~\cite{EXOT-2017-01,EXOT-2019-01,EXOT-2018-05}, an additional charged lepton~\cite{EXOT-2018-32}, or $b$-tagging of the jets~\cite{EXOT-2022-33,EXOT-2016-33,EXOT-2019-03,CMS-EXO-20-008,CMS-EXO-19-004,CMS-EXO-17-027} by ATLAS and CMS\@.
These searches allowed ATLAS  to cover a wider region of dijet mass from \qty{100}{\GeV} up to multiple \unit{\TeV} than previous searches by ATLAS, albeit with limited acceptance due to the additional requirements and, consequently, less sensitivity than the TLA result.

This paper presents an updated TLA search for dijet resonances in the mass range \qtyrange{375}{1800}{\GeV} using the full statistical power of the sample recorded by ATLAS during Run~2.
Superseding the prior ATLAS TLA result~\cite{EXOT-2016-20}, this new search incorporates four times the integrated luminosity, an improved first-level trigger strategy to extend the search to lower dijet masses, an improved jet calibration, and more sophisticated statistical analysis techniques.
%

%
%

% End of text imported from the .//01intro.tex input file

% The next lines are included from the .//02detector.tex input file
%

\section{ATLAS detector}
\label{sec:detector}

\newcommand{\AtlasCoordFootnote}{%
ATLAS uses a right-handed coordinate system with its origin at the nominal interaction point (IP)
in the centre of the detector and the \(z\)-axis along the beam pipe.
The \(x\)-axis points from the IP to the centre of the LHC ring,
and the \(y\)-axis points upwards.
Cylindrical coordinates \((r,\phi)\) are used in the transverse plane,
\(\phi\) being the azimuthal angle around the \(z\)-axis.
The pseudorapidity is defined in terms of the polar angle \(\theta\) as \(\eta = -\ln \tan(\theta/2)\).
Angular distance is measured in units of \(\Delta R \equiv \sqrt{(\Delta\eta)^{2} + (\Delta\phi)^{2}}\).}

The ATLAS detector~\cite{PERF-2007-01} at the LHC covers nearly the entire solid angle around the collision point.\footnote{\AtlasCoordFootnote}
It consists of an inner tracking detector surrounded by a thin superconducting solenoid, electromagnetic and hadron calorimeters, and a muon spectrometer incorporating three large superconducting air-core toroidal magnets.
The inner-detector system (ID) is in a \qty{2}{\tesla} axial magnetic field and provides charged-particle tracking in the range \(|\eta| < 2.5\).
The high-granularity silicon pixel detector covers the vertex region and typically provides four measurements per track, the first hit normally being in the insertable B-layer (IBL) installed before Run~2~\cite{ATLAS-TDR-19,PIX-2018-001}.
It is followed by the silicon microstrip tracker (SCT), which usually provides eight measurements per track.
These silicon detectors are complemented by the transition radiation tracker (TRT),
which enables radially extended track reconstruction up to \(|\eta| = 2.0\).
The TRT also provides electron identification information based on the fraction of hits (typically 30 in total) above a higher energy-deposit threshold corresponding to transition radiation.

The calorimeter system covers the pseudorapidity range \(|\eta| < 4.9\).
Within the region \(|\eta|< 3.2\), electromagnetic calorimetry is provided by barrel and endcap high-granularity lead/liquid-argon (LAr) calorimeters, with an additional thin LAr presampler covering \(|\eta| < 1.8\) to correct for energy loss in material upstream of the calorimeters.
Hadron calorimetry is provided by the steel/scintillator-tile calorimeter, segmented into three barrel structures within \(|\eta| < 1.7\), and two copper/LAr hadron endcap calorimeters.
The solid angle coverage is completed with forward copper/LAr and tungsten/LAr calorimeter modules optimised for electromagnetic and hadronic energy measurements respectively.

The muon spectrometer comprises separate trigger and high-precision tracking chambers measuring the deflection of muons in a magnetic field generated by the superconducting air-core toroidal magnets.

Collision events of interest are selected by a first-level (\LevelOne) trigger system implemented in custom hardware, followed by selections made by algorithms implemented in software in a high-level trigger (\HLT)~\cite{TRIG-2016-01}.
The L1 trigger accepts events from the \qty{40}{\MHz} bunch crossings at a rate below \qty{100}{\kHz} and features a topological processor, \LevelOneTopo~\cite{TRIG-2019-02}, allowing selections on multi-body quantities such as the angular distance between two jets.
The \HLT reconstructs events accepted by the \LevelOne to record an average of \qty{1.2}{\kHz} of full events with an average size of \qty{1}{\mega\byte} to permanent storage in a “main” stream~\cite{TRIG-2019-04}.
Specialised streams record additional data for low-\pT flavour physics analysis, monitoring, calibration, and other uses.
These streams use dedicated triggers, non-standard reconstruction paths, or readout of only subsets of the detector around regions of interest.
For TLA, a dedicated stream stores the \HLT jet information~\cite{TRIG-2012-01,TRIG-2016-01} for all events accepted by certain L1 trigger selections, regardless of the \HLT decision, while discarding the raw detector data found in the main stream.
The information recorded for each \HLT jet consisted of the four-momentum and calorimeter variables providing information about the quality and structure of each jet~\cite{ATLAS-CONF-2015-029}.
No tracking measurements were recorded for TLA\@.
The TLA stream recorded events with a rate of up to \qty{27}{\kHz} at a mean \qty{6.5}{\kilo\byte} per event~\cite{TRIG-2019-04}.
An extensive software suite~\cite{SOFT-2022-02} is used in the detector operations, trigger and data acquisition systems of the experiment, and in simulation, reconstruction, and analysis.

%
%
%

% End of text imported from the .//02detector.tex input file

% The next lines are included from the .//03samples.tex input file
%

\section{Data and simulation samples}
\label{sec:samples}

\begin{figure}[h!]
\centering
\includegraphics[width=0.8\linewidth]{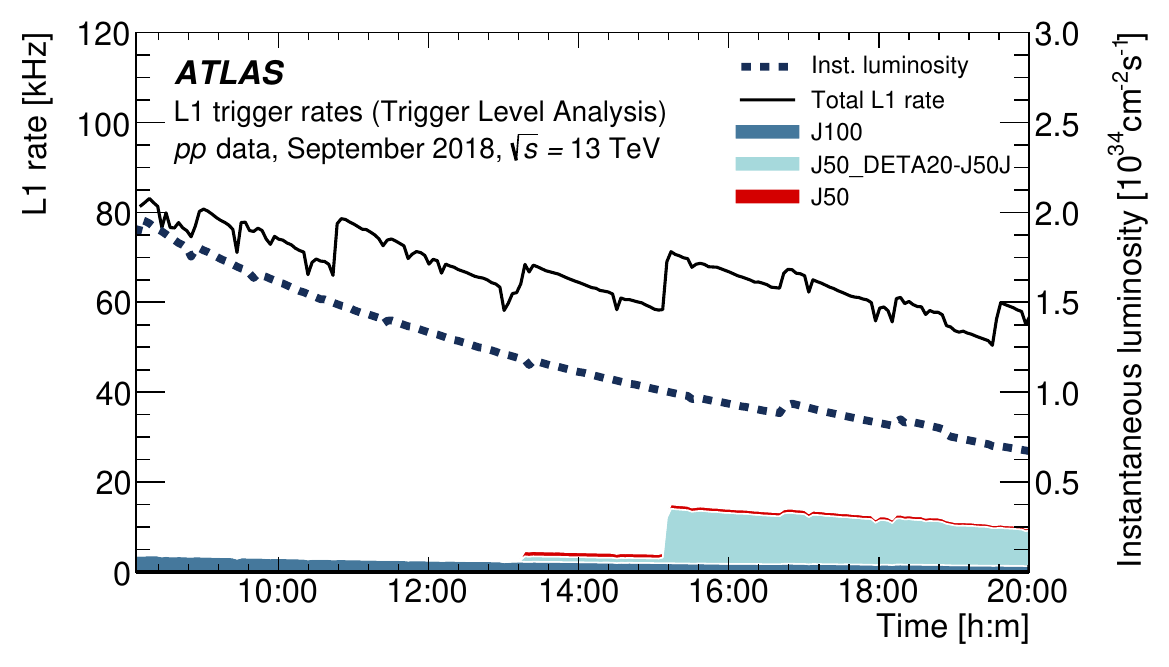}
\caption{
Trigger rate as a function of time for the total L1 trigger and the L1 triggers feeding the TLA stream.
The single LHC fill shown occurred in September 2018.
While the J100 L1 item (dark blue histogram) collected events for TLA throughout the fill, the single-jet J50 and dijet J50\_DETA20-J50J triggers (light blue and red histograms) activated as the LHC instantaneous luminosity (dashed line) and total L1 trigger rate (solid line) declined during the fill.
}
\label{fig:trigops}
\end{figure}

Data were selected by several \LevelOne triggers and processed by the TLA stream for offline analysis as follows.
In each collision, the L1 trigger identified \LevelOne jets from $\Delta \eta \times \Delta \phi = 0.2 \times 0.2$ electromagnetic and hadronic calorimeter segments with a sliding-window algorithm.
Events containing at least one \LevelOne jet within $|\eta|<3.1$ with sufficient transverse momentum were recorded in the TLA stream.
From 2016–2018, this requirement was $\pT>\qty{100}{\GeV}$, which corresponds to the lowest-threshold, unprescaled single-jet \LevelOne trigger for most of that period.
Starting in 2017, additional prescaled triggers with lower thresholds were included at the end of LHC fills as declining instantaneous luminosity during a fill lowered the actual trigger rate of the initial \LevelOne menu well below its capability, yielding a peak rate of \qty{27}{\kHz}.
In 2017, a single end-of-fill trigger required an \LevelOne jet with $\pT>\qty{50}{\GeV}$.
In 2018, this was expanded to a luminosity-dependent combination of triggers, from which events were selected for analysis with either of two triggers.
The first required a single \LevelOne jet with $\pT>\qty{50}{\GeV}$.
The second, making use of the \LevelOneTopo processor, required two \LevelOne jets satisfying $\pT>\qty{50}{\GeV}$ for the leading jet, $\pT>\qty{15}{\GeV}$ for the sub-leading jet, and  $|\yStar|<1$, where $|\yStar|\equiv|\Delta\eta/2|$ is half of the absolute rapidity difference between the two jets.
No data recorded in 2015 are considered, as only a small amount of luminosity was recorded with the TLA stream that year, requiring an \LevelOne jet with $\pT>\qty{75}{\GeV}$.
That sample is fully analysed in the prior ATLAS dijet TLA~\cite{EXOT-2016-20}.
\cref{fig:trigops} shows the typical rate for the individual TLA streams during an LHC fill.

In addition to the data used for the search, collision events recorded with the main stream and reconstructed offline were used to calibrate the jets used for the search and to validate the statistical methods applied to the TLA stream.
Events in the main stream provide a superset of the information available for events in the TLA stream and are thus suitable to derive a calibration for the latter.
The two streams were triggered independently, with the decision to record an event in one stream unaffected by the decision to record an event in the other.
The main stream was not used to construct the search sample.
Main stream events were recorded with a suite of unprescaled and prescaled single-jet triggers with various \pT thresholds as low as \qty{45}{\GeV}.
The unprescaled single-jet trigger with the lowest threshold required $\pT>\qty{420}{\GeV}$ at the \HLT selection.

All data used from the TLA and main streams were recorded in stable beam conditions with all relevant detector systems operating normally~\cite{DAPR-2018-01}.
Events are discarded if occurring during brief periods associated with calorimeter data corruption or transient LAr calorimeter noise~\cite{LARG-2013-01}.
The number of $pp$ interactions per bunch crossing, referred to as pile-up, spanned 10--70 interactions per bunch crossing ($\mu$), with an average of $\mu=25$ in 2016, $\mu=38$ in 2017, and $\mu=36$ in 2018.
Pile-up both from the same bunch crossing (``in-time'') and preceding bunch crossings (``out-of-time'') affects the energy reconstruction in the calorimeters and is corrected for in a dedicated step of the jet calibration.

\begin{figure}[h]
\centering
\includegraphics[width=0.8\linewidth]{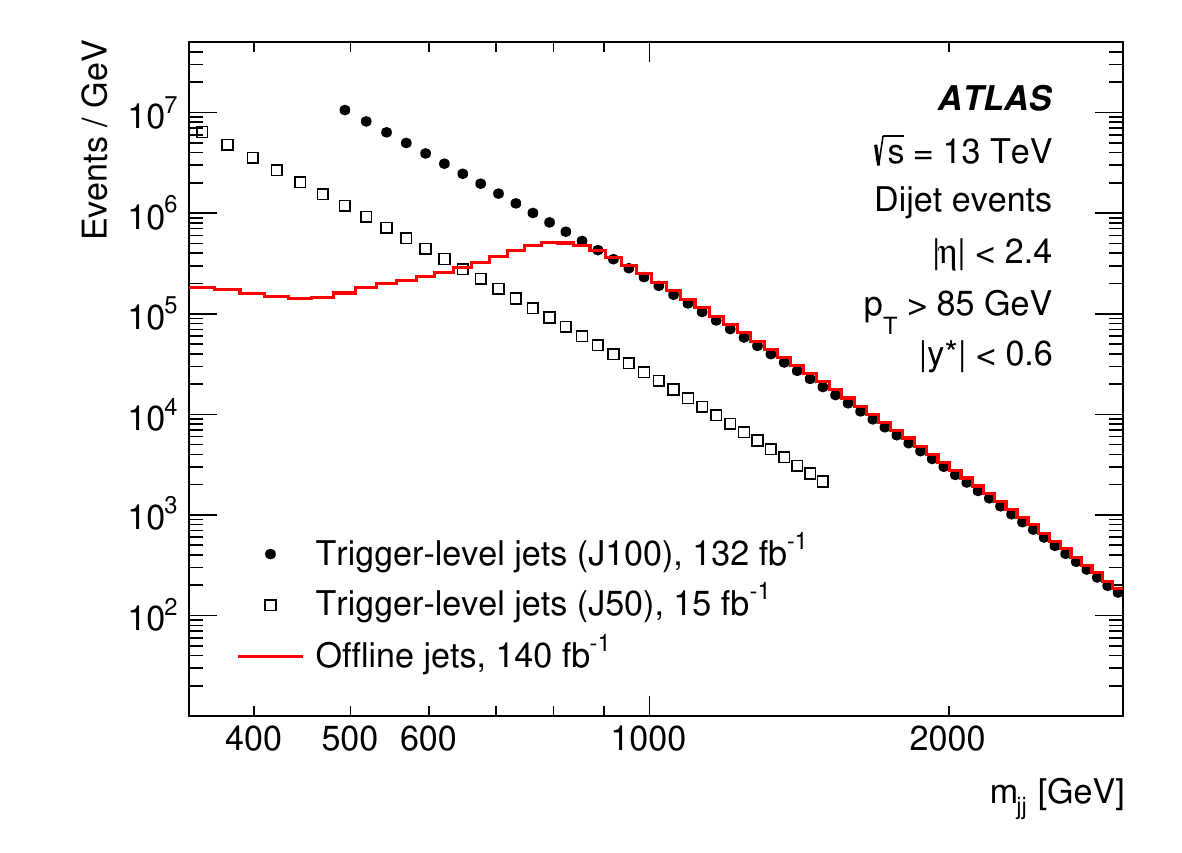}
\caption{Distributions of dijet invariant mass \mjj for events collected via the TLA stream and reconstructed in the \HLT, combining events selected at \LevelOne by the $\pT > \qty{100}{\GeV}$ trigger (filled circles) and the prescaled single-jet and dijet $\pT > \qty{50}{\GeV}$ triggers (open squares), showing the increase in events recorded at lower values of dijet mass for this analysis compared with events collected via the main stream and reconstructed offline (histogram).
For this illustrative comparison, each sample must satisfy the analysis selection described in \cref{sec:selection}, except for rejection of the tile gap region, and includes events with at least two \onlineJetsAdjective or offline jets, each with $\pT>\qty{85}{\GeV}$ and $|\eta|<2.4$, with half a rapidity difference smaller than 0.6.
The final sample used for the search is discussed and shown separately in \cref{sec:selection}.
}
\label{fig:scoutingyield}
\end{figure}

After trigger and data-quality requirements, this search uses \intLumiJHundred of data collected from 2016 to 2018 with the TLA stream via the $\pT>\qty{100}{\GeV}$ \LevelOne trigger (``J100'') and \intLumiJFifty recorded with the $\pT>\qty{50}{\GeV}$ \LevelOne end-of-fill single-jet and dijet triggers (``J50'').
These two datasets overlap since J50-triggered events with a jet of sufficiently high \pT also fulfil the requirements of the unprescaled J100 trigger.
\cref{fig:scoutingyield} compares the samples collected by the TLA and main streams as a function of dijet invariant mass.

Relative to the sample collected with the main stream using both prescaled and unprescaled triggers, the J50 TLA sample provides up to $\sim 30$ times greater numbers of events per mass bin and the J100 TLA sample provides up to $\sim 75$ times greater numbers of events.

Events from simulation of multijet processes are used to derive calibrations of jet energy measurements and evaluate the effects of certain systematic uncertainties, described below.
These events were simulated with \PythiaDijet~\cite{Sjostrand:2014zea}, the A14~\cite{ATL-PHYS-PUB-2014-021} set of tuned parameters for the underlying event, and the leading-order \NNPDF~\cite{Ball:2012cx} parton distribution functions (PDFs).
The renormalisation and factorisation scales were set to the average \pT~of the two leading jets.
In order to model the effect of pile-up, the hard-scatter event was overlaid with simulated minimum-bias events generated with \PythiaPileup~\cite{Sjostrand:2007gs} using the leading-order \NNPDF PDFs and the A3 set of tuned parameters~\cite{ATL-PHYS-PUB-2016-017}.
The events were reweighted such that the resulting distribution of average number of interactions per bunch crossing matches that observed in data.
Detector effects were simulated using \GeantFour~\cite{Agostinelli:2002hh} within the ATLAS software infrastructure~\cite{SOFT-2022-02,SOFT-2010-01}.
Separate samples were constructed for each year of data taking to reflect the changes in luminosity profile and detector conditions.

Additional spin-1 \Zprime bosons often arise in models of phenomena beyond the Standard Model (BSM), such as in those involving symmetry breaking of extended gauge theories, Kaluza--Klein towers arising from extra dimensions, and in many others.
As a benchmark for these and other resonances, a leptophobic \Zprime model~\cite{Abercrombie:2015wmb} was simulated at leading order, with matrix elements calculated in \MadGraph~\cite{Alwall:2014hca} using the leading-order \NNPDF PDFs and parton showering performed in \PythiaZprime with \EvtGen~\cite{Lange:2001uf} and the A14 set of tuned parameters.
Detector effects including the effects of pile-up were simulated as for the sample of multijet processes using the ATLAS simulation.
The \Zprime model assumes axial-vector couplings to all SM quarks with one coupling value common to all quark flavours.
The \Zprime simulation did not model interference with the SM, such as with the \Zboson boson.
The model includes a coupling of the \Zprime to a dark matter candidate, but the simulation adopted a scenario~\cite{Chala:2015ama,Abercrombie:2015wmb} where its decays into dark matter are negligible.
In this case, the dijet production rate and resonance width depend, at leading order, only on the coupling to quarks, $g_q$, and the mass of the resonance, $m_{\Zprime}$.
The production rate scales quadratically with $g_q$.

%
%
%
%

% End of text imported from the .//03samples.tex input file

% The next lines are included from the .//05calibration.tex input file
%

\section{Jet reconstruction and calibration}
\label{sec:calibration}

Jets were reconstructed by the \HLT (``\onlineJetsAdjective jets'') solely from topological clusters of signals in contiguous calorimeter cells~\cite{PERF-2014-07} using the \antiKt algorithm~\cite{Cacciari:2008gp,Fastjet} with a radius parameter of $R=0.4$.
Each cluster is assigned an energy equal to the sum of the associated cell energies. These signals are calibrated at the electromagnetic (EM) scale, accounting correctly for the energy deposited by EM showers but not by hadronic showers. Consequently, the resulting jets formed from such topological clusters (EMTopo) require further calibration.
For this search, only \onlineJetsAdjective jets are available.
The analysis discards jets with $\pT<\qty{85}{\GeV}$ or $|\eta|>2.4$ to minimise the contribution of pile-up jets and to achieve the best possible jet resolution.
Events that contain jets induced by calorimeter noise, beam-induced background, or cosmic rays were rejected using the ``BadLoose'' criteria identical to Ref.~\cite{ATLAS-CONF-2015-029} but with no requirement on the track-based charged fraction, which was not recorded for \triggerLevel jets but has a negligible effect after the requirement on the rapidity difference described in \cref{sec:selection}.

A calibration procedure was performed for \onlineJetsAdjective jets during the analysis to correct their energy and direction to those of particle-level jets.
This calibration made use of two additional types of jet reconstruction.
For main stream data, used in the final calibration step, an additional collection of jets was reconstructed from the raw detector data (``offline jets'') with an equivalent \antiKt $R=0.4$ algorithm and kinematic selections, calibrated using the standard ATLAS techniques~\cite{PERF-2016-04}.
For simulated data, \antiKt $R=0.4$ particle-level jets were reconstructed from stable particles with a lifetime greater than \qty{10}{\pico\second} (``truth jets'') and serve as the reference for the \onlineJetsAdjective jet calibration.

\begin{figure}[h!]
\centering
\includegraphics[width=1.\linewidth]{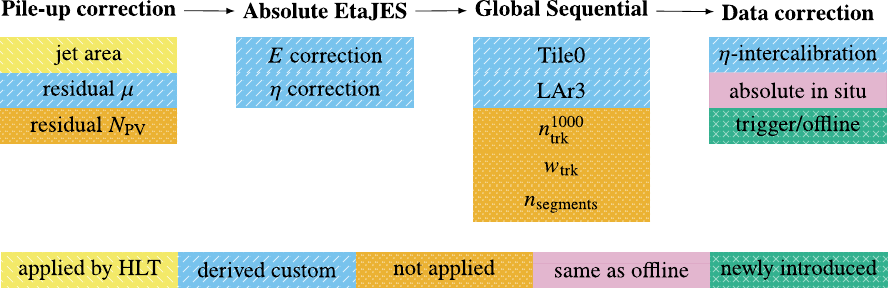}
\caption{
Stages and substeps of the \onlineJetsAdjective jet calibration and how they differ from the offline jet calibration.
Following the jet area correction performed by the \HLT during data collection (yellow hatched), this analysis uses dedicated calibration steps (blue hatched) correcting for pile-up and differing energy response of hadronic signals throughout the detector.
Steps of the offline calibration employing tracking are omitted due to the lack of tracking in the TLA stream (orange hatched).
Trigger jets in data are further corrected with a dedicated $\eta$-intercalibration while the same absolute in situ calibration as derived for offline jets is applied (purple hatched).
This is followed by a new trigger/offline step (green hatched) to correct for remaining differences between offline and trigger jets.
}
\label{fig:calibration}
\end{figure}

The calibration of \onlineJetsAdjective jets closely followed the usual calibration procedure for offline jets~\cite{PERF-2016-04}.
The differences between \onlineJetsAdjective and offline calibrations, summarised in \cref{fig:calibration} and described in the following, arise from the differences in jet reconstruction and the information available at the trigger level.
A final calibration step, the \onlineJetsAdjective/offline correction, was performed to align \onlineJetsAdjective jets to the offline jet energy scale.
This allowed the treatment of \onlineJetsAdjective jets to adopt identical uncertainties to those derived during the offline jet calibration, except for flavour uncertainties, whose treatment differs from that for offline jets due to the lack of tracking information and is discussed in this section and \cref{sec:uncertainties}.
All steps up to and including the Global Sequential correction are applied to jets both in simulation and in data, while all subsequent steps are applied only to jets in data.

The high statistical precision of the TLA stream, with a sensitivity to signal amplitudes of the order of \(10^{-4}\) to \(10^{-3}\) relative to the background contribution, poses strict smoothness requirements on the jet calibration.
Throughout the calibration chain, special emphasis was placed on keeping the jet response a smooth function of jet \pT without local features to avoid introducing peaking features into the dijet mass distribution.

\paragraph{Pile-up correction}
The first step in the calibration sequence was the jet-area-based pile-up correction, to remove additional energy from pile-up collisions.
The jet area $A$, as a measure for the susceptibility of a given jet to additional energy from pile-up collisions, was measured for each jet by ghost-association~\cite{Cacciari:2008gn, Cacciari:2007fd}.
The contribution of pile-up in the event was estimated from the average pile-up density \mbox{$\rho=\text{median}\left\{\pT(j)/A(j)\right\}$} of the ratio of transverse momentum \pT and area $A$ of all jets $j$ in the event.
The initial transverse momentum \ptinit was corrected to
\begin{align*}
\ptarea = \ptinit - \rho A,
\end{align*}
using the product of $\rho$ and $A$ as an estimate for the contribution from pile-up to the momentum of each individual jet.
This correction was performed within the \HLT.
Therefore the \onlineJetsAdjective jets entering the TLA stream were calibrated to this scale and all subsequent calibration steps were derived and applied relative to it.

The offline jet pile-up calibration~\cite{PERF-2016-04} procedure foresees a residual correction
\begin{align*}
\ptresidual = \ptarea - \frac{\partial\pT}{\partial\NPV} (\NPV - 1) - \frac{\partial\pT}{\partial\avgMu} \avgMu,
\end{align*}
based on the number of vertices with at least two associated tracks with \(\pT > \qty{500}{\MeV}\) (``primary vertices'') \NPV in the event and the mean number of interactions per bunch crossing \avgMu.
The former is affected only by in-time pile-up while the latter is affected by both in-time and out-of-time pile-up.
Since vertex information is not recorded for the TLA stream, a calibration based on \NPV is not possible; hence only the $\avgMu$ term of the residual pile-up calibration was applied to \onlineJetsAdjective.
In this and all subsequent simulation-based calibration steps, \onlineJetsAdjective and truth jets were geometrically matched by their angular distance requirement $\Delta R<0.3$, and truth (\onlineJetsAdjective) jets were required to be isolated from all other truth (\onlineJetsAdjective) jets in the event by $\Delta R>1.0$ ($\Delta R>0.6$).
The factor $\partial\pT/\partial\avgMu$ was determined by linear fits of $\ptarea - \pTtruth$ versus $\avgMu$ for \onlineJetsAdjective jets with $60 < \pTtruth < \qty{80}{\GeV}$ in simulated dijet events.
This was done independently for each $|\etaTruth|<4.0$ interval of width $0.1$.

\begin{figure}[h!]
\centering
\subfloat[]{\includegraphics[width=0.8\linewidth, ]{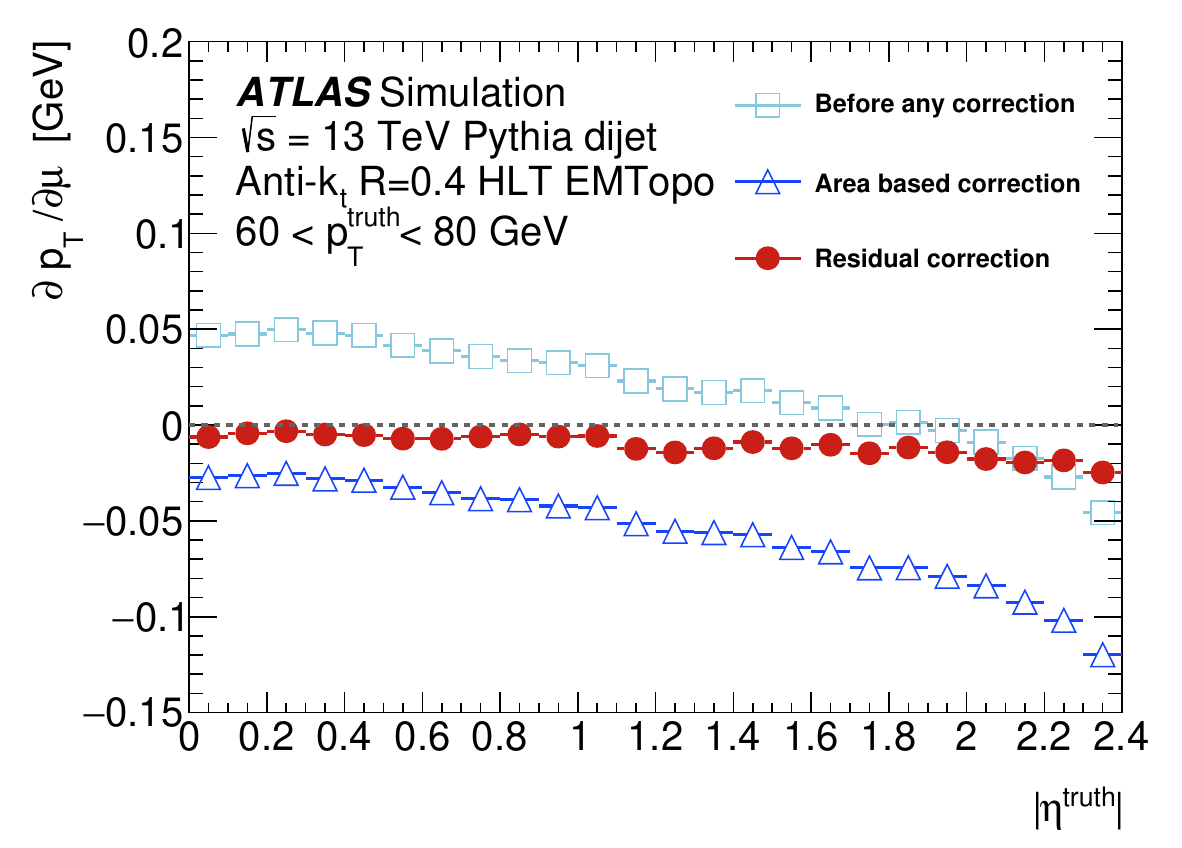}\label{fig:pileup_mu}}\\
\subfloat[]{\includegraphics[width=0.8\linewidth, ]{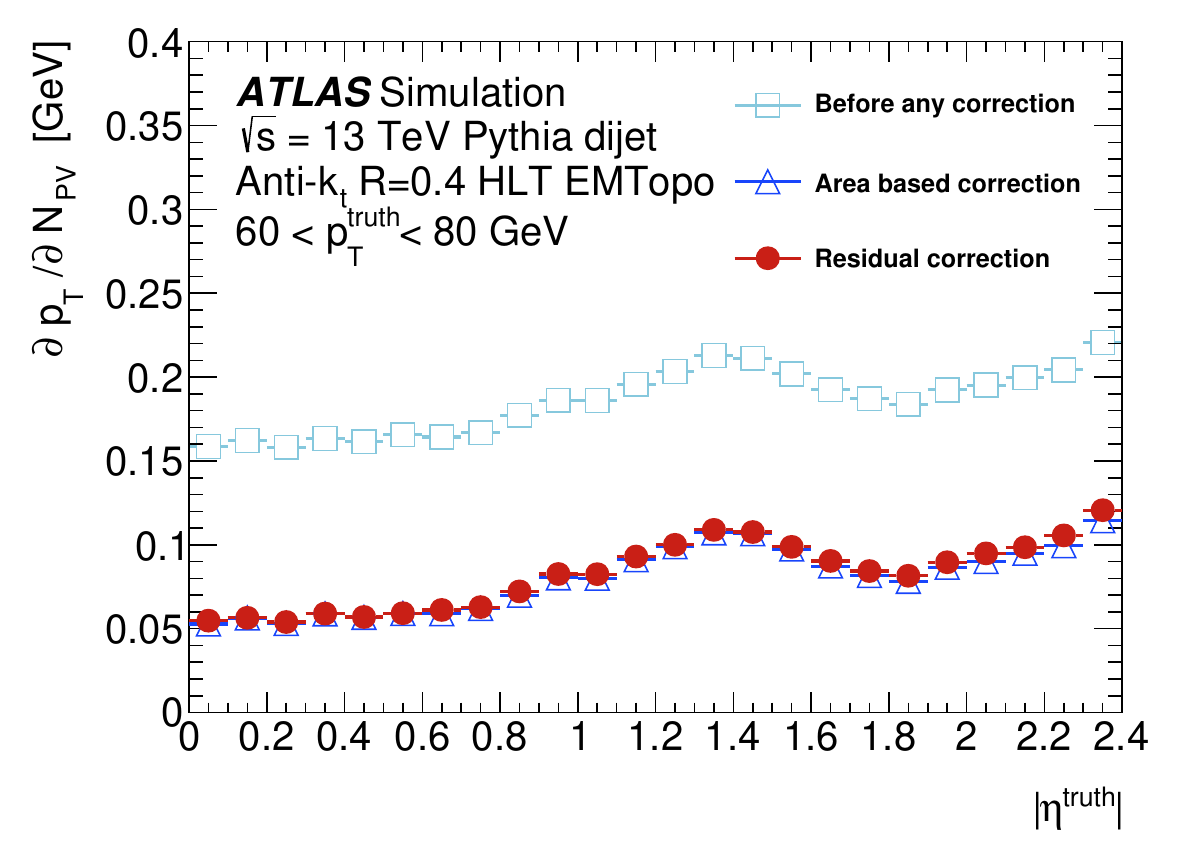}\label{fig:pileup_npv}}
\caption{
Dependence of the reconstructed \onlineJetsAdjective jet \pT on \protect\subref{fig:pileup_mu} the average number of interactions per bunch crossings \avgMu and \protect\subref{fig:pileup_npv} the number of primary vertices \NPV before (open squares) and after the two steps of pile-up calibration described in \cref{sec:calibration}, the area based correction (open triangles), and residual correction (filled circles) in simulated dijet events, as a function of the pseudorapidity $|\eta^\text{truth}|$ of the corresponding truth jet.
}
\label{fig:pileup}
\end{figure}

\cref{fig:pileup_mu} shows that this residual step significantly decreases the dependence $\partial\pT/\partial\avgMu$ of \onlineJetsAdjective jet \pT with respect to \avgMu.
\cref{fig:pileup_npv} shows that the $\partial\pT/\partial\NPV$ dependence is unaffected by this correction.
This constitutes a significant improvement on the prior ATLAS dijet TLA~\cite{EXOT-2016-20}, where only the jet-area-based correction was applied.

\paragraph{Absolute EtaJES correction}
Subsequently, a simulation-based absolute calibration was applied to correct the jet energy $E$ scale (JES) to that of truth jets.
The average jet energy response after the pile-up correction was measured as $\mathcal{R}_E=\frac{\Eonline}{\Etruth}$ in bins of \Etruth and \etaDetector.\footnote{
\etaDetector is the jet pseudorapidity pointing from the geometric centre of the detector.
Its use reduces ambiguity as to which region of the detector is measuring the jet.
}
The statistical mode of each response distribution was estimated as the mean of a fit of a Gaussian distribution to its core.
Correction factors were extracted by fitting the resulting modes with the functional form
\begin{align*}
\mathcal{F}_\text{JES}(\Etruth)=\sum_{i=0}^{N_{\textrm{max}}}a_i(\text{ln}~\Etruth)^i,
\end{align*}
where $a_i$ is a free parameter.
For each fit the functional order $N_\text{max}$ is chosen from the interval \(\lbrace 1 \, .. \, 6 \rbrace\) such that it results in the minimal $\chi^2$ of the fit~\cite{PERF-2011-03}.
A numerical inversion procedure was applied to obtain the factors as a function of \onlineJetsAdjective rather than truth quantities $\mathcal{F}_\text{JES}(\Eonline)$, such that they can be applied to \onlineJetsAdjective jets.

\begin{figure}[h!]
\centering
\subfloat[]{\includegraphics[width=0.8\linewidth]{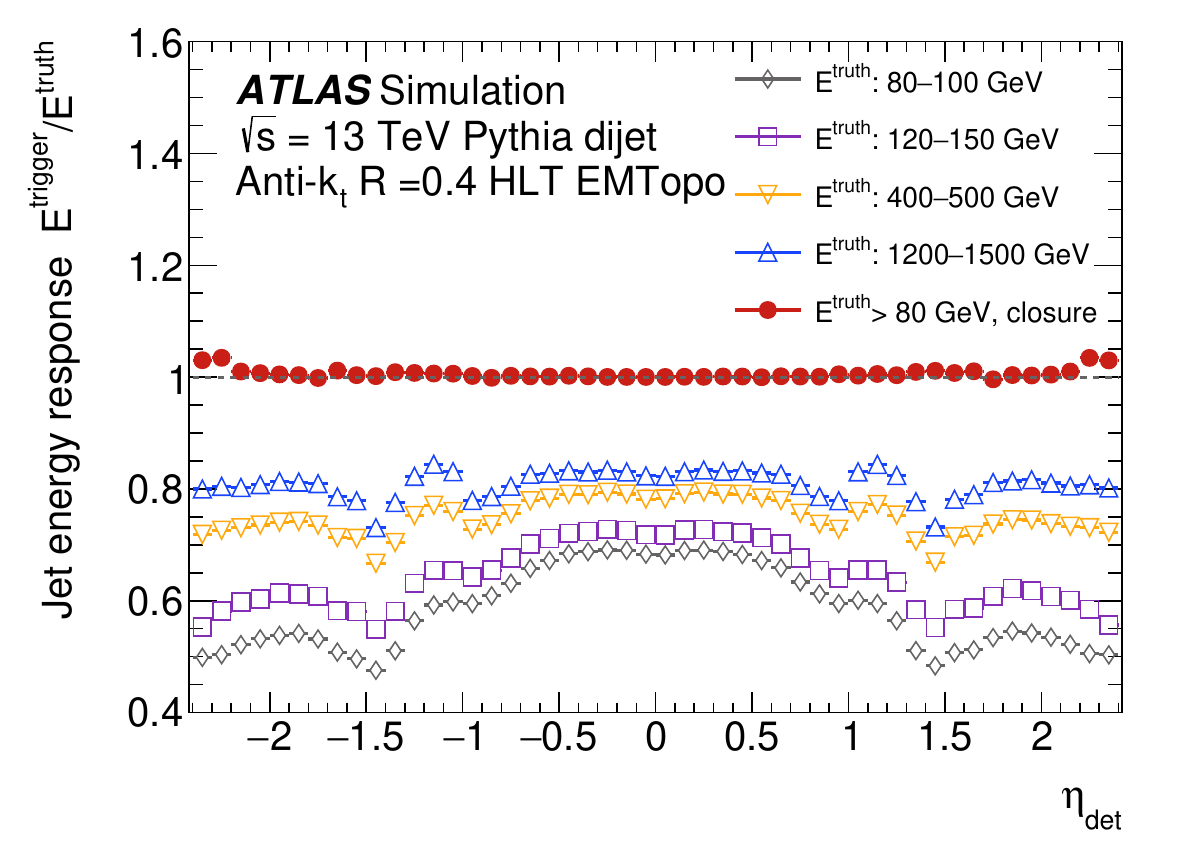}\label{fig:etajes_response}}\\
\subfloat[]{\includegraphics[width=0.8\linewidth]{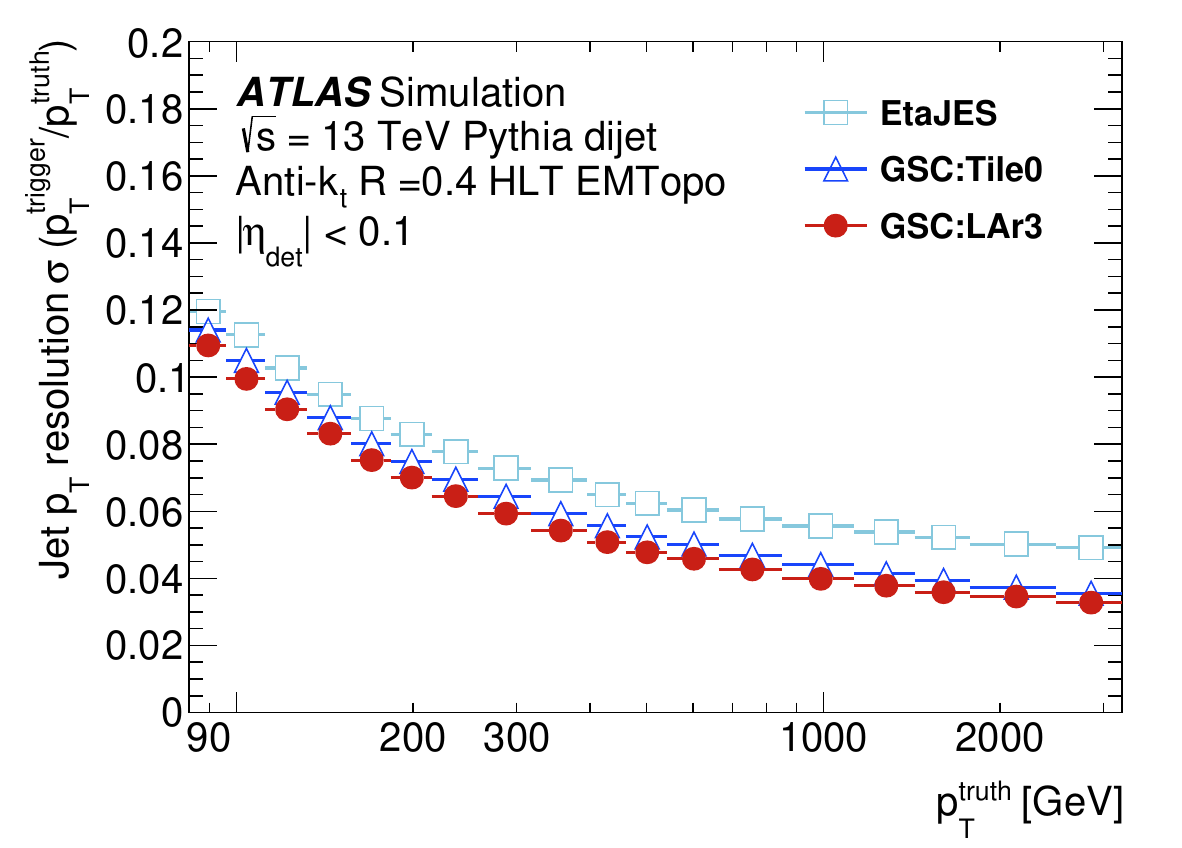}\label{fig:etajes_resolution}}
\caption{
\protect\subref{fig:etajes_response} \OnlineJetsAdjective jet energy response, $E^{\textrm{\onlineJetsAdjective}}/\Etruth$, in bins of detector pseudorapidity \etaDetector of the truth jet \pT, $\pT^{\textrm{truth}}$, before applying the EtaJES calibration step (open markers in ranges of truth jet energy from \qtyrange{80}{1500}{\GeV}) and closure after applying the correction (filled circles). \protect\subref{fig:etajes_resolution} Resolution $\sigma(\pT^{\textrm{\onlineJetsAdjective}}/\pT^{\textrm{truth}})$ of \onlineJetsAdjective jet transverse momentum \pT as a function of truth jet \pT after the EtaJES calibration step (open squares) and after each of the calorimeter-based GSC steps, Tile0 (open triangles) and LAr3 (filled circles) in the central detector $\eta$ bin $|\etaDetector|<0.1$.
}
\label{fig:etajes}
\end{figure}

\cref{fig:etajes_response} shows that the value of the response before applying the EtaJES correction is highly dependent on the pseudorapidity \etaDetector and can change by up to $50\%$.
This is partly due to varying instrumentation across different detector regions, which can result in a bias in the direction of reconstructed jets, especially the pseudorapidity $\eta$, towards better-instrumented regions of the detector.
Hence, an additional correction to the pseudorapidity was applied to compensate for this effect.
It was derived in the same way as the energy correction, but using the difference $\Delta\eta=\etaOnline - \etaTruth$ rather than the response $\mathcal{R}_E$.

Closure is defined as the difference between the response to unity after the correction is applied.
\cref{fig:etajes_response} shows that the EtaJES achieves a closure of $5\%$ in the forward region (and significantly below that in the central region) for jets with $E>\qty{80}{\GeV}$.

\paragraph{Global Sequential correction}
Up to this point both energy and direction of reconstructed \onlineJetsAdjective jets are calibrated on average.
A Global Sequential correction (GSC) was then applied to improve the response of individual populations of jets, e.g. jets that differ in  the fraction of energy deposited in different calorimeter components, while leaving the overall response unchanged.
As a result, the overall resolution improves, as measured by the width of Gaussian distributions fit to the cores of the response.
ATLAS typically applies a GSC~\cite{PERF-2016-04} based on five variables:
\begin{itemize}
\item $f_\text{Tile0}$: The fraction of jet energy measured in the first layer of the hadronic tile calorimeter;
\item $f_\text{LAr3}$: The fraction of jet energy measured in the third layer of the electromagnetic Liquid Argon calorimeter;
\item $n_\text{trk}^{1000}$: The number of $\pT>\qty{1}{\GeV}$ tracks associated with the jet;
\item $w_\text{trk}$: The \pT-weighted transverse distance of tracks to the jet axis;
\item $n_\text{segments}$: The \pT-weighted transverse distance of muon spectrometer segments to the jet axis.
\end{itemize}
Due to their dependence on tracking information, the last three of these variables were not available to the TLA stream.
Consequently, only the first two steps of the normal sequence, involving the variables $f_\text{Tile0}$ and $f_\text{LAr3}$, were performed for the \onlineJetsAdjective jet calibration.
One GSC variable after the other, the response $\mathcal{R}_{\pT}$ distribution of \onlineJetsAdjective jets with respect to matched truth jets was constructed in bins of \pTtruth, \etaDetector, and the GSC variable and extracted as the mean of a fit to a Gaussian distribution.
For each $\eta$ bin, a 2-dimensional (2-D) Gaussian kernel smoothing in \pTtruth and the GSC variable was performed on the resulting average response.
The final calibration factors as a function of \onlineJetsAdjective jet \pT and $\eta$ were determined by numerical inversion.
This procedure was performed for each GSC variable sequentially, i.e.~the response of the second variable was constructed after applying the correction based on the first variable.
\cref{fig:etajes_resolution} illustrates both steps of the GSC improve the \onlineJetsAdjective jet resolution.

The subsequent tracking-based steps of the offline GSC sequence are intended to correct for the differences between quark- and gluon- initiated jets.
These latter steps use variables ($n_\text{trk}^{1000}$, $w_\text{trk}$, $n_\text{segments}$) available only from offline reconstruction.
They typically produce corrections that are smaller than the prior steps in the offline sequence.
Despite lacking the track-based steps, the \onlineJetsAdjective GSC described here achieves similar resolution to the offline GSC, with $10\%$ at \qty{100}{\GeV} and 4~\% at \qty{2}{\TeV}~\cite{JETM-2018-05}. %
Flavour uncertainties for \onlineJetsAdjective jets were re-derived to account for the absence of the tracking-based steps and found to be a factor of two larger than for offline jets, as discussed in \cref{sec:uncertainties}.

\paragraph{Data correction}
In addition to the three Monte-Carlo derived calibration steps described so far, a data-based calibration was applied, with the goal of correcting for residual differences in the energy scale in data that are not modelled in the simulation.
While the Monte-Carlo derived calibrations rely on simulated particle-level jets as reference, the data corrections rely on reconstructed objects with well-calibrated response as reference to calibrate objects with less well-understood response, taking advantage of the overall transverse momentum balance in an event.
For offline jets, this correction comprises two parts: the $\eta$-intercalibration and the absolute in situ calibration.
The former balances jets in the forward detector regions with those in the better instrumented central region while the latter balances jets in hadronic recoil events with a photon or leptonically-decaying \Zboson boson.
For \onlineJetsAdjective jets, an additional step, the \onlineJetsAdjective/offline correction, was derived.

The $\eta$-intercalibration is designed to accommodate for the effect that the response of central jets is usually better constrained than that of forward jets.
The ratio of data / simulation response is constructed in bins of $\eta$ and \pT and a 2-D smoothing is applied.
A jet-vertex-tagger (JVT)~\cite{PERF-2014-03} requirement of $>0.59$ is applied based on the JVT score of $\Delta R<0.3$ matched offline jets.
To better constrain the smoothness, for \onlineJetsAdjective jets a 7-parameter poly-log functional form fit
\begin{align*}
\mathcal{F}_\text{intercal}(\pT) = \sum_{i=-3}^{3}a_i \log^i(\pT)
\end{align*}
was used instead of the 2-D Gaussian kernel used for the offline jet in situ calibration.
Apart from this, the procedure of deriving the \(\eta\)-intercalibration here does not differ from that for offline jets.
The resulting distribution is applied as a calibration curve for jets with $|\eta|>0.8$.

For offline jets, an absolute in situ calibration was obtained from the \pT balance of events with jets recoiling against well-measured objects in data.
Forward jets were corrected based on them recoiling against central jets in dijet events while central jets were corrected based on recoils with photons in $\gamma+$jets events or leptons in $\Zboson(\rightarrow\text{ leptons})+$jets events.
The previous calibration steps bring the \onlineJetsAdjective jets to nearly identical scales as  offline jets, which allowed the \onlineJetsAdjective jet calibration to use the same in situ calibration factors as offline jets.

An additional calibration step correcting \onlineJetsAdjective jets to the scale of offline jets was applied to account for any residual differences between the two calibrations.
This calibration was derived independently for each year (2016, 2017, 2018) and the two TLA data samples (J50, J100) accounting for the different pile-up conditions of each year of data taking, and the fact that the J50 L1 trigger was mostly active at the end of LHC fills (again with differing pile-up conditions).
To store both \onlineJetsAdjective and offline jets for an event, the \HLT must have accepted the event, initiating the offline jet reconstruction.
Different single-jet \HLT triggers were used depending on the $\pTavg=(\pT^\textrm{lead}+\pT^\textrm{sublead})/2$, the average transverse momentum of \pT-leading and sub-leading offline jets $\Delta R<0.2$ matched to \onlineJetsAdjective jets, ensuring the triggers are fully efficient.
Event-by-event prescale factors were taken into account to correct the population of jets for bias from triggers whose varying prescale factors led to differing distributions of instantaneous luminosity.

A response $\mathcal{R}=\pTonline/\pToffline$ was derived as the mean of fits in data to a Gaussian distribution in bins of $\pTavg$.
The motivation for deriving and validating the \onlineJetsAdjective/offline calibration in bins of \pTavg rather than single-jet \pT was to account for the varying flavour composition of the two leading jets per event such that the correction factor is valid for the whole dijet system.
A numerical inversion, followed by a 2-D Gaussian kernel smoothing, was performed to determine the final calibration factor in bins of \onlineJetsAdjective jet \pTavg and $\eta$.

\begin{figure}[h!]
\centering
\includegraphics[width=0.8\linewidth]{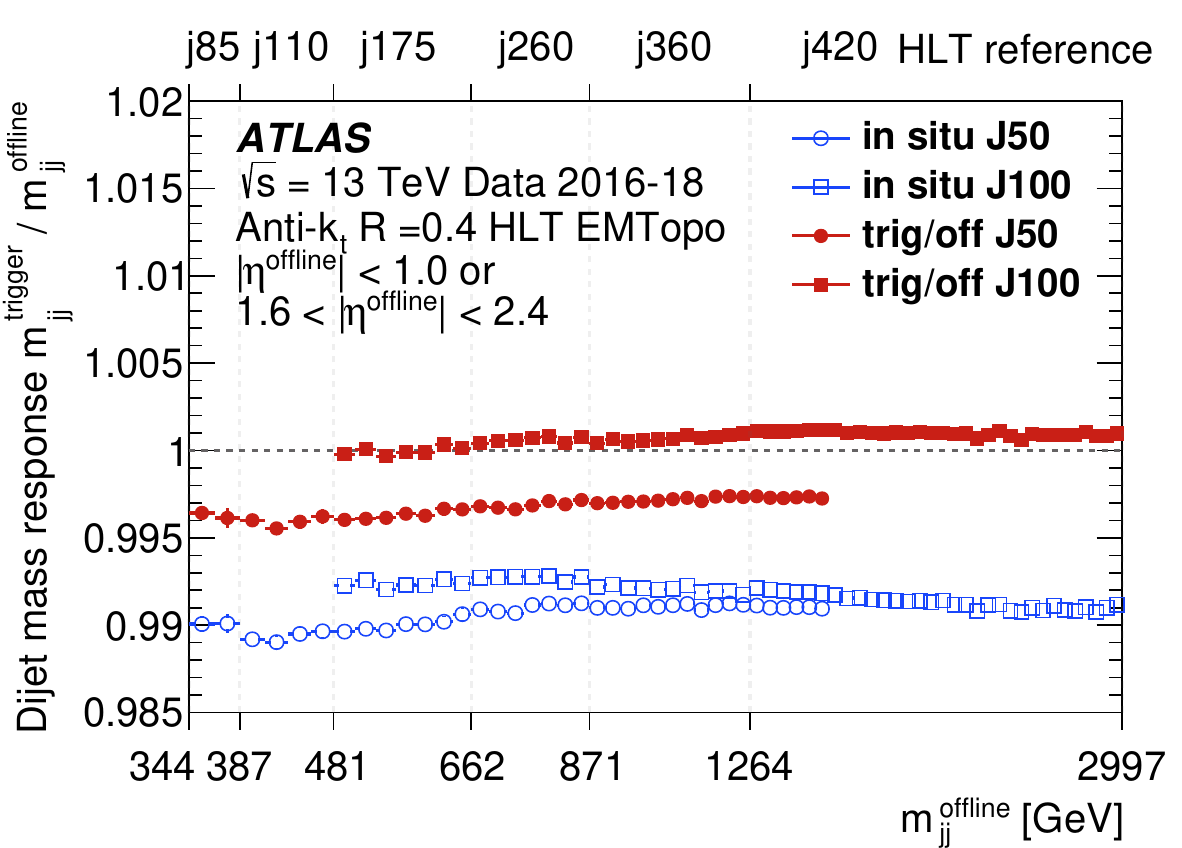}
\caption{
Improvement in the ratio of \onlineJetsAdjective to offline dijet mass, $\mjj^{\textrm{\onlineJetsAdjective}}/\mjj^{\textrm{offline}}$, for the J50 (circles) and J100 (squares) samples before (open markers after the in situ calibration step) and after applying the \onlineJetsAdjective/offline calibration (filled markers) for all years of data taking combined. For offline jets, different leading jet $\pT>X$~\unit{\GeV} thresholds (abbreviated as j$X$) are required which are fully efficient for a given $m_{jj}^\text{offline}$ region.
Each region is exclusively populated by one trigger based on the offline jet $\pT$.
Events where either leading or sub-leading jet fall into the region $1.0<|\eta^\text{offline}|<1.6$ are excluded to avoid jets whose calibration performance is not sufficient, as discussed in \cref{sec:selection}.
}
\label{fig:onlineofflineMjjAll}
\end{figure}

A \onlineJetsAdjective/offline \pT closure of the order of $1\%$ is observed inclusively in $\eta$.
After the correction, it improves to be within $0.5\%$ over the full \pTavg range.
The closure in individual $\eta$ regions before (after) the correction is within $2\%$ ($1\%$).
The final statistical interpretation of the analysis is performed on the invariant mass of the dijet system \mjj.
Hence, the \mjj response of \onlineJetsAdjective relative to offline jets, \cref{fig:onlineofflineMjjAll}, was measured as a final calibration check.
Similarly to the \pT closure, the \mjj closure improves from $1\%$ to $0.5\%$ after applying the correction.
The differences in closure between J50 and J100 arise from the difference in pile-up conditions, with the J50 selection activated exclusively during lower instantaneous luminosity.
The remaining non-closure is approximately constant in \mjj and small in comparison to the dominant jet energy scale uncertainties discussed in \cref{sec:uncertainties}.

\clearpage

%
%
%
%

% End of text imported from the .//05calibration.tex input file

% The next lines are included from the .//06kinematic.tex input file
\section{Event selection}
\label{sec:selection}

For the search and the data-driven background estimate, events from the TLA stream were selected that contain at least two \onlineJetsAdjective jets, each with $\pT>\qty{85}{\GeV}$ and $|\eta|<2.4$.
The two highest \pT jets satisfy $|\yStar|<0.6$.
This requirement suppresses the background from SM multijet processes, which grows with $\yStar$ due to $t$-channel poles at leading order.
This value was chosen to optimise the Poisson signal significance ($S/\sqrt{B}$) in simulated data for the \Zprime signals of mass \qtyrange{350}{1000}{\GeV}.

After these selections, events with either jet having $1.0 < |\eta|<1.6$ were discarded.
These jets overlap with a narrow ``tile gap'' region between hadronic calorimeters that is instrumented only by a thin set of gap and cryostat scintillators~\cite{TCAL-2021-01}.
In this region, the jet calibration procedure described in the previous section is unable to reach the performance demanded by the statistical uncertainties in the TLA sample.

Events meeting these requirements were further divided into ``J100'' and ``J50'' signal regions by the L1 trigger that collected them.
For J50-triggered data, events with calibrated \mjj below \qty{344}{\GeV} were discarded to remove data sculpted by the kinematic requirements of the L1 trigger.
For J100-triggered data, the minimum allowed $\mjj$ was \qty{481}{\GeV}.
These thresholds were chosen from dedicated measurements of the trigger efficiencies as a function of data-taking period, jet $\pT$ and dijet \mjj, varying the $\yStar$ requirement from $0.3$ to no requirement at all, to ensure that these triggers reached full efficiency throughout the mass regions used for the search.

\begin{figure}[h]
\centering
\includegraphics[width=0.8\linewidth,]{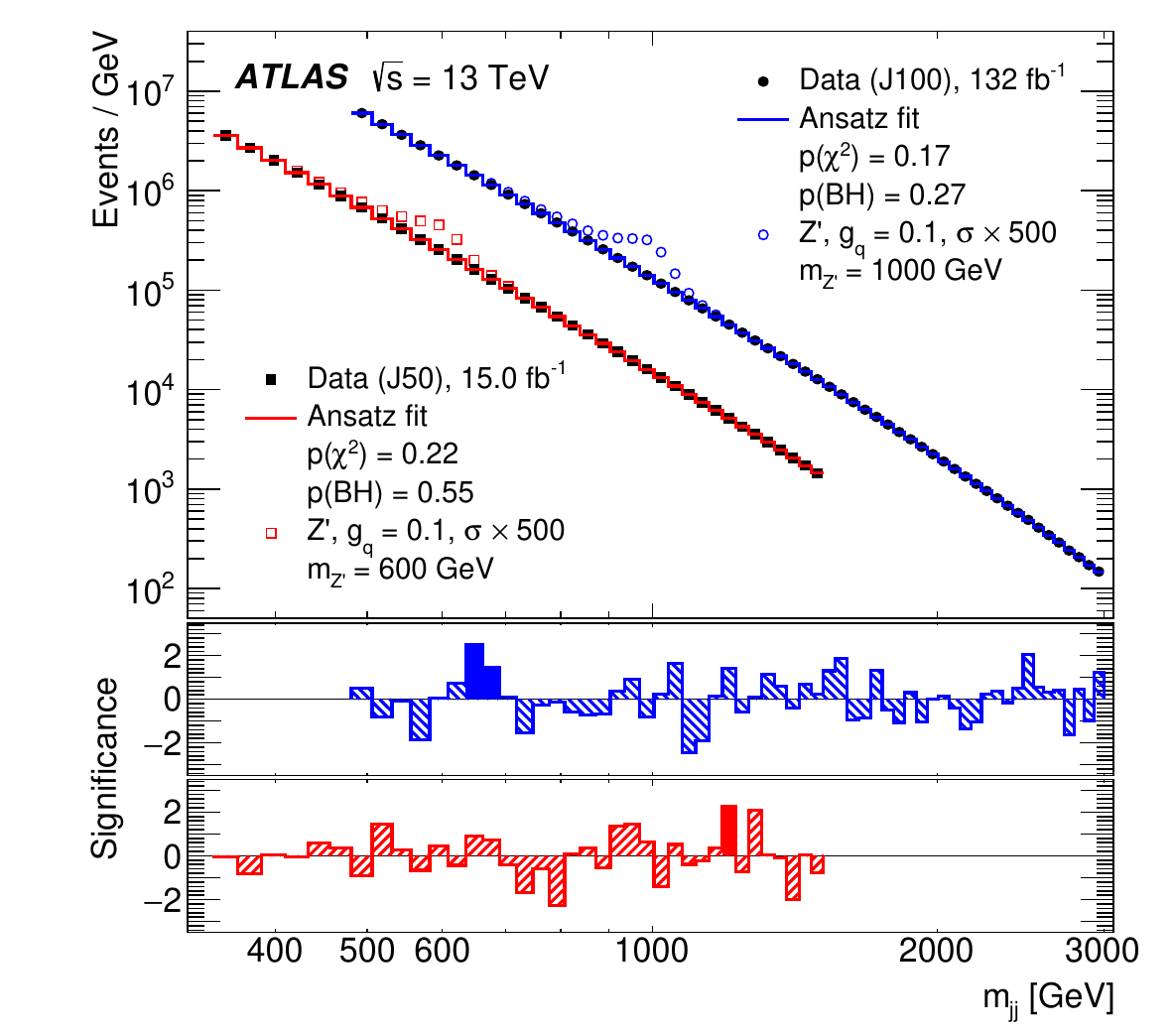}
\caption{Observed \mjj distributions for the J50 (filled squares) and J100 (filled circles) signal regions, using the variable-width bins and analysis selections described in the text.
In the upper panel, the histograms indicate the background estimates from each ansatz fit, integrated over each bin.
Predictions for two \Zprime signals are shown above the fit (open markers), normalised to 500 times the predicted cross-section for a quark coupling $g_q=0.1$ for visibility.
The lower panels show the significance of the differences between data and background estimates in units of the statistical uncertainties in the numbers of events in each bin.
The regions identified as the most significant by \BumpHunter are indicated in the bottom panel as contiguous solid-filled bins.
The \(p\)-value for the fit \(\chi^2\) and global value from \BumpHunter are indicated in the legends for each of the two signal regions.
}
\label{fig:searchmjj}
\end{figure}

\cref{fig:searchmjj} shows the \mjj distributions for the two signal regions after these selections, in variable-width bins which correspond to the calibrated \onlineJetsAdjective jet \mjj resolution in simulation.

%
%

% End of text imported from the .//06kinematic.tex input file

% The next lines are included from the .//07statana.tex input file
%

\section{Background estimate}
\label{sec:background}

The SM background in the search is dominated by non-peaking multijet processes convolved with a smooth detector response.
The estimate of this background began with an ansatz
\begin{equation}
\label{eq:backgroundfit}
f(x)=p_{1}(1-x)^{p_{2}}x^{\sum_{i=3}^N p_{i} \log^{(i-3)}(x)} \big( 1 + \sum_{j=1}^{n_\textrm{syst}} \theta_j \sum_{k=1}^5 c_{jk} x^k \big),
\end{equation}
where $x \equiv \mjj/\sqrt{s}$ and $N$ is a fixed number of shape parameters.
Dijet mass distributions were found to be well-described by expressions similar to \cref{eq:backgroundfit} ($\theta_j=0$) in prior dijet searches with lower statistical precision~\cite{CDF:2009DijetSearch,EXOT-2010-01,CMS-EXO-11-015,EXOT-2010-07,EXOT-2013-11,EXOT-2019-03,CMS-EXO-19-012}.
In \cref{eq:backgroundfit}, systematic uncertainties in the background shape due to the jet calibration are modelled via nuisance parameters $\theta_j$ and modulating factors $c_{jk}$ derived from simulation as described in \cref{sec:uncertainties}.
The \mjj data in \cref{fig:searchmjj} in each signal region were divided into \qty{1}{\GeV} fixed-size bins and fit with \cref{eq:backgroundfit} using a \RooFit+\RooStats-based~\cite{RooFit,moneta2011roostatsproject} framework to obtain the parameters $p_{i}$ and nuisance parameters $\theta_j$.

Fits to this ansatz were validated separately for each of the two signal regions.
As constructing a sample of multijet processes with equivalent statistical uncertainties to \cref{fig:searchmjj} was computationally impractical with the ATLAS detector simulation, the fit procedure for fixed $N$ was validated on pseudodata.
This pseudodata was generated by applying Poisson fluctuations matching the statistical uncertainties in \cref{fig:searchmjj} to fits of both subsamples and the entirety of the data in \cref{fig:searchmjj} with \cref{eq:backgroundfit} with $N_0=N+1$ parameters.
A \chisq probability larger than 1\% was required for the $N_0$ fit to ensure that the pseudodata accurately models the shape of the \mjj distribution in data.
In case of a localised excess prohibiting the \chisq threshold being reached, the excess region would have been masked from the $N_0$ fit to ensure that pseudodata corresponding to a background-only \mjj distribution could be generated.

Fits to the pseudodata confirm the \chisq goodness-of-fit follows the eponymous distribution, which was used to assign the quoted \chisq probabilities with the number of degrees of freedom given by the difference between the number of bins and the number of unconstrained parameters in the fit.

The fit for fixed $N$ was validated by performing signal-plus-background fits with freely floating signal amplitude on the generated $N_0$ pseudodata, using a set of the signal masses and shapes tested in \cref{sec:results}.
As the pseudodata is generated to behave like the background-only distribution, the expectation of the signal amplitude extracted in these fits is zero.
Any deviation from this indicates a bias in the fit and is thus labelled spurious signal.
For each signal hypothesis it was validated on the ensemble of fits to pseudodata that the spurious signal is Gaussian-distributed with a mean of the order of the standard deviation.
Additionally, $N_0$ pseudodata with signals injected were used to verify that the background component of the fit is not significantly affected by the presence of signal and that the extracted signal amplitude corresponds to the injected amplitude within uncertainties.
Notably, the machine precision of the bin-wise evaluation of the negative log-likelihood in \RooFit~ was found insufficient for the unprecedented numbers of events shown in \cref{fig:searchmjj}; the Gaussian limits of the Poissonian terms in the likelihood were used to avoid this limitation.

From the previously validated fit strategies, an $F$-test~\cite{ATL-PHYS-PUB-2020-028} was performed to determine the value of $N$ for which further increasing the fit flexibility does not significantly improve the \chisq goodness-of-fit. %
This test was performed on the data in both signal regions independently with a threshold of \(p(F)>0.05\).
It resulted in a choice of \(N=6\) in both signal regions.%

\section{Uncertainties}
\label{sec:uncertainties}

\begin{figure}[h!]
\centering
\subfloat[]{\includegraphics[width=.49\linewidth, trim={3 0 -3 0}]{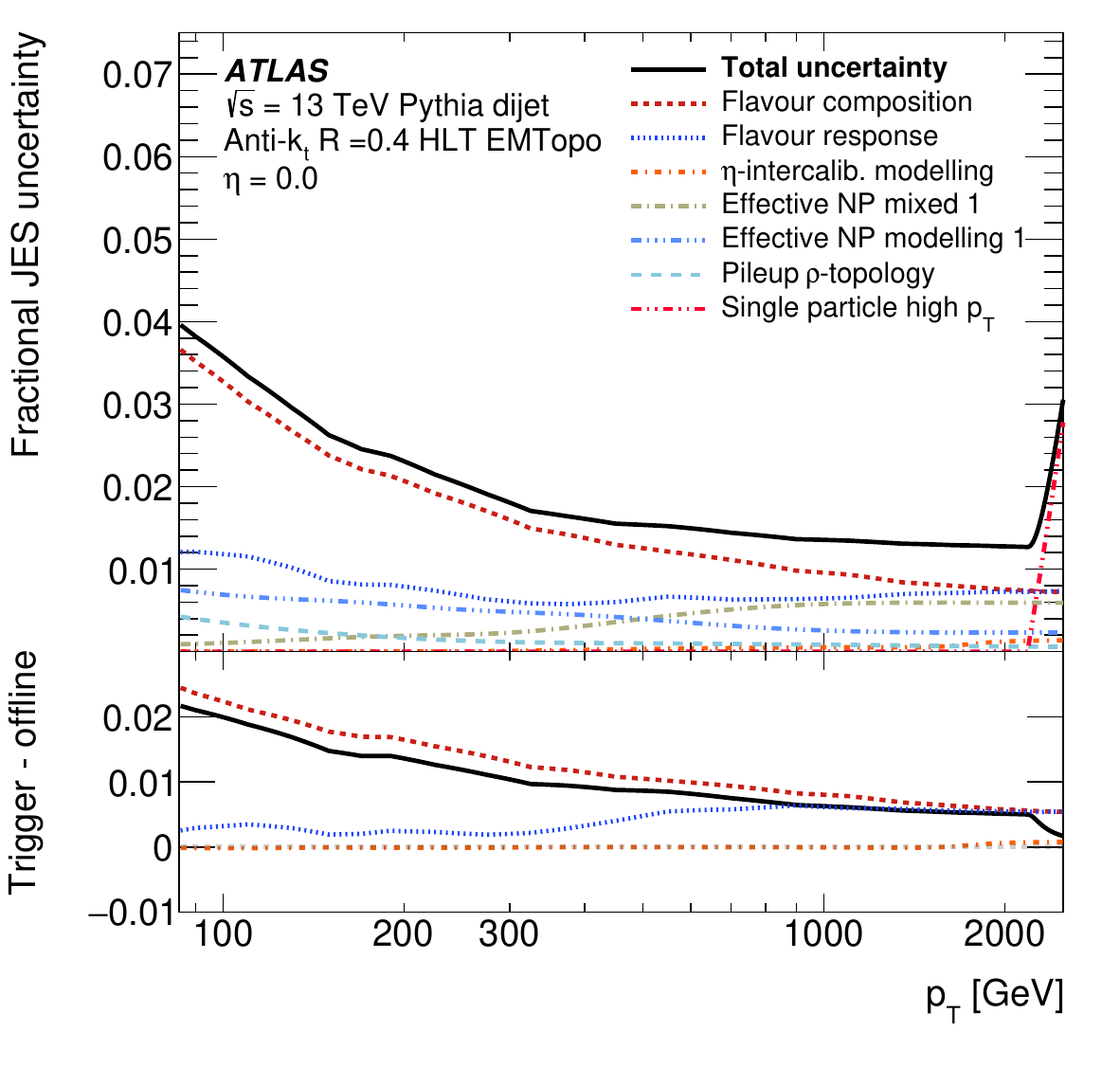}\label{fig:uncerts_pt}}
\hfill
\subfloat[]{\includegraphics[width=.49\linewidth, trim={-7 0 7 0}]{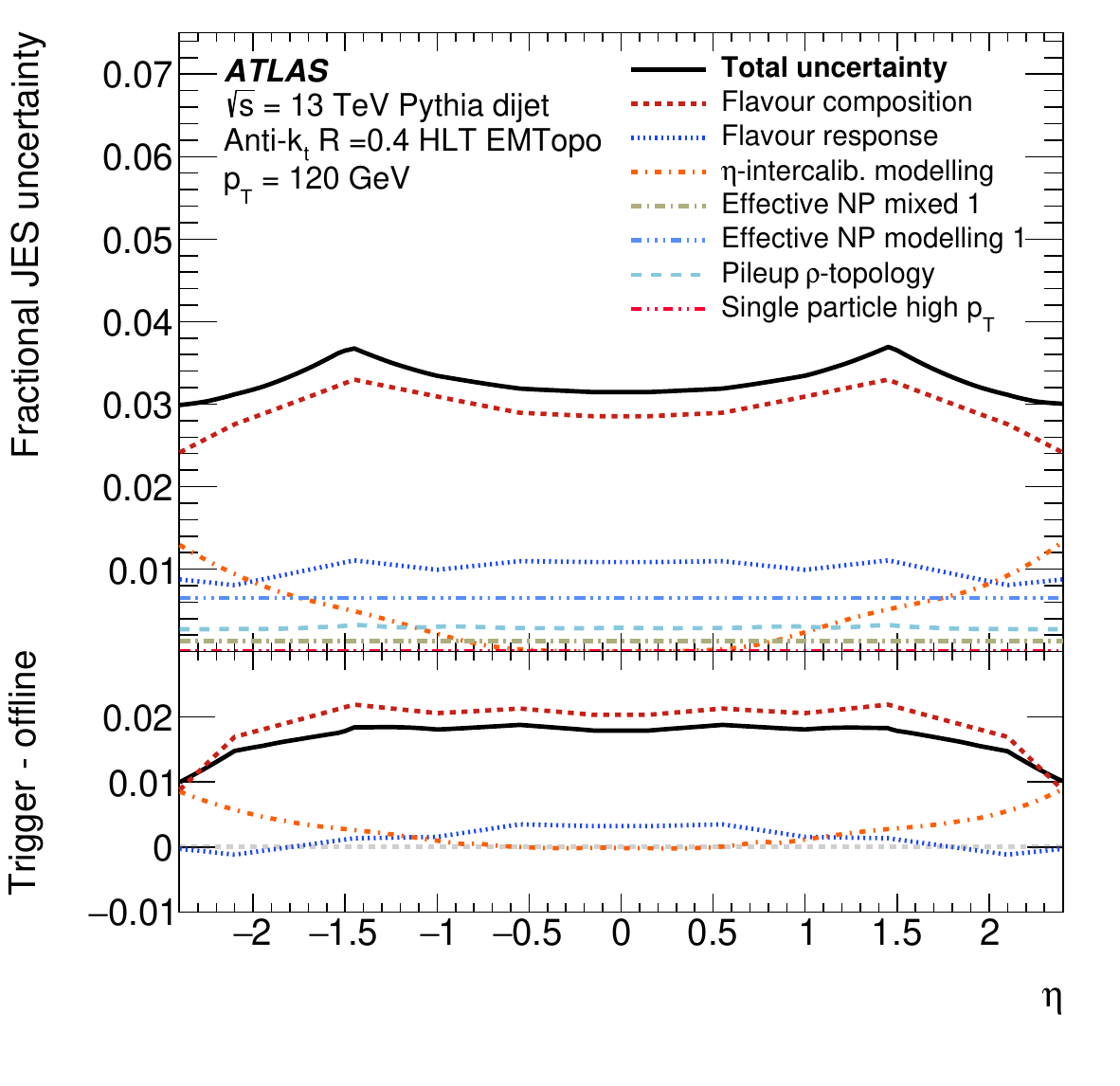}\label{fig:uncerts_eta}}
\caption{
Total systematic uncertainty in the energy scale of \onlineJetsAdjective jets (solid lines) and the most significant of 40 components considered, as a function of jet \protect\subref{fig:uncerts_pt} $p_\text{T}$ and \protect\subref{fig:uncerts_eta} $\eta$.
The bottom panel shows the absolute differences between \onlineJetsAdjective jet and offline jet uncertainties for the total uncertainty and the three components that were re-derived for \onlineJetsAdjective jets.
Uncertainties in the flavour composition of jets (uppermost dashed line) are the largest contribution to the total uncertainty and the source of the largest difference with the offline calibration uncertainty, arising due to the absence of tracking information used in the offline GSC to correct for different responses to quark- and gluon-initiated jets.
}
\label{fig:uncerts}
\end{figure}

Systematic uncertainties in the jet energy scale and resolution were considered for both the signal and the background contributions.
They were evaluated on the simulated \Zprime and dijet samples mentioned in \cref{sec:samples}, respectively, by independently varying 40 nuisance parameters up and down within their uncertainties on a jet-by-jet basis.
The jet energy scale uncertainties with the highest impact on the signal template \Zprime fits are shown in \cref{fig:uncerts}.
All uncertainties except for the flavour composition, flavour response, and $\eta$-intercalibration modelling uncertainty are identical to those derived for offline jets.

The flavour uncertainties were re-derived as in Ref.~\cite{JETM-2018-05},
\begin{align*}
\sigma^\text{flavour}_\text{response} &= f_g \sigma_{R_g},\\
\sigma^\text{flavour}_\text{composition} &= \frac{\sigma_{f_g} |R_q - R_g|}{f_g  R_g + (1 - f_g) R_q },
\end{align*}
where $R_q$ and $R_g$ are the fraction of light quarks and gluons that were estimated from \Pythia dijet simulated events.
The gluon response modelling uncertainty $\sigma_{R_g}$ was obtained as the maximum difference between the gluon response in \Pythia, \Herwig~\cite{Bahr:2008pv,Bellm:2015jjp}, and \Sherpa~\cite{Hoeche:2009rj,Bothmann:2019yzt} simulated events. %
The gluon fraction $f_g$ was set to a constant value of $50\%$ with a conservative estimate on its uncertainty of $\sigma_{f_g}=100\% f_g$.
The flavour uncertainties are the dominant jet uncertainties and mainly arise due to the lack of tracking information which, for the offline jet calibration, is key to aligning the response of quark- and gluon-initiated jets~\cite{JETM-2018-05}.

To parameterise the impact of the uncertainties on the signal, a double-sided Crystal Ball function~\cite{Oreglia:1980cs,HIGG-2014-04} was fit to the nominal and systematically varied \Zprime \mjj distributions.
The impact of each nuisance parameter on the mean and the width of the Gaussian core of the function is assigned as systematic uncertainty both for \Zprime and Gaussian-shaped resonance signals.
The dominant sources of uncertainty in the mean of the signal is the uncertainty in the jet flavour composition ranging from \qtyrange{8}{20}{\GeV} with rising signal mass.
It is followed by the jet flavour response and the in situ uncertainties both ranging approximately from \qtyrange{2}{12}{\GeV}.
The uncertainty in the width of the signal is small with all nuisance parameters having a sub-\unit{\GeV} impact.

For the systematic uncertainties in the background, the ratios of the systematic up and down variations of the \mjj distributions to the nominal distribution were symmetrised and fit with a fifth degree polynomial.
The resulting coefficients $c_{jk}$, $k=1...5$ for each systematic uncertainty $j$ were used to parameterise the shape of this uncertainty in the background ansatz in \cref{eq:backgroundfit}, treating it as fully correlated across \mjj.
The amplitude for each is a nuisance parameter $\theta_j$ with normal constraint in the likelihood.

The systematic uncertainties in the signal and background contributions were treated as uncorrelated.

The amount of spurious signal determined for each signal hypothesis and mass was considered as an additional systematic uncertainty by adding a separate signal component of identical shape and a Gaussian-constrained amplitude to the signal-plus-background fits.
This uncertainty was found to be up to the order of half the statistical uncertainty in the signal amplitude for most \Zprime and narrow Gaussian-shaped resonance signals and up to \(170\%\) of the statistical uncertainty for wide resonances.
The extreme cases are largely driven by the inclusion of the systematic uncertainties in background shape in the ansatz fit, which increases the fit flexibility and the spurious signal.
The resulting impact on the sensitive signal cross-section ranges from being negligible to up to an \(80\%\) increase, depending on signal shape and mass.

The uncertainties originating from the total collected luminosity of the samples were below \(1\%\) and have a negligible effect on the presented results~\cite{DAPR-2021-01}.
No other normalisation uncertainties in the \Zprime cross-section for a given coupling constant \(g_q\) is applied for comparability with other results probing \Zprime dijet resonances.

\section{Results and exclusion limits}
\label{sec:results}

The \mjj distributions obtained from the J50 and J100 signal regions are shown in \cref{fig:searchmjj}, overlaid with the corresponding background estimates from the $N=6$ ansatz fits including background systematic variations.
The residuals for each fit are shown in the lower panels.
The observed distributions are well described by the fits.
The fits yield a \chisq $p$-value of 0.22 in the J50 signal region and 0.17 in the J100 region.

The \BumpHunter algorithm~\cite{Aaltonen:2008vt, Choudalakis:2011bh} was used to further assess localised excesses in either \mjj distribution.
The algorithm scans all possible contiguous mass intervals across the binned distribution, from a width of one bin to a width of half of the distribution.
For each interval, it computes the probability of a Poisson fluctuation of the fitted background estimate at least as large as that observed.
The algorithm identifies the bins spanning \qtyrange{1186}{1225}{\GeV} for the J50 distribution and \qtyrange{635}{690}{\GeV} for the J100 distribution, indicated by solid-filled bins in the residual panels of the figure, as the most significant discrepancies in each signal region.
The global \(p\)-values, determined from a large number of pseudoexperiments drawn from the background estimate, are 0.55 for the J50 signal region and 0.27 for the J100 signal region.
Thus, this yields no evidence of a localised contribution to the mass distributions in excess of the background.

%
%
%
%
%
%

% End of text imported from the .//07statana.tex input file

% The next lines are included from the .//08limits.tex input file
%

The data were used to constrain the \Zprime model discussed in \cref{sec:samples} and the cross-section times acceptance times branching ratio to two jets, $\sigma\times A\times\text{BR}$, of a hypothetical signal that produces a Gaussian-shaped contribution to the observed \mjj distribution.

Samples of \Zprime events were simulated for \MZprime values of \num{200}, \num{350}, \num{450}, \num{650}, \num{1000}, and \num{2000} \unit{\GeV} and at coupling values $g_q=0.2$ and lower.
Before parton shower effects, the intrinsic width of this \Zprime is 1.6\%--1.9\% depending on \MZprime for $g_q=0.2$ and at the lower coupling values reached by the search ranges from 0.1\% of its mass for a mass of \qty{850}{\GeV} and $g_q=0.04$ to 0.6\% for a mass of \qty{1.8}{\TeV} and $g_q=0.11$.
The decrease in width with decreasing coupling was negligible compared to the detector \mjj resolution and therefore neglected.
The selection acceptance in the J50 signal region for a signal with $g_q=0.1$ is 12\% for a mass of \qty{375}{\GeV} and in the J100 signal region is 17\% for a mass of \qty{600}{\GeV} and 32\% for a mass of \qty{2.0}{\TeV}.
Each simulated sample was analysed identically to data to obtain an \mjj line-shape.
Line-shapes for additional values of \MZprime were generated by linear interpolation of the parameters of a double-sided Crystal Ball function fit to these distributions. %

Limits were also calculated for hypothetical signals with a Gaussian-shaped \mjj distribution at mean mass $m_G$ and three different widths from 5\% to 15\% of the mean mass.

To derive the constraints on these two types of signals from the \mjj distribution, a Frequentist method~\cite{Cowan2011} was applied to the data and simulation of signals at a series of discrete masses to set 95\% $\textrm{CL}_\textrm{S}$~\cite{Read:2002hq} upper limits on the cross-section times acceptance times branching ratio to jets for the signals.

Only the signal region with the stronger expected limit was used for each signal mass.
The J100 sample, with higher statistical power, constrained masses above \qty{600}{\GeV}, while lower masses, more affected by fit uncertainties due to the minimum \mjj requirement, were constrained using the J50 sample.

When evaluating the likelihood to derive limits, the parameters of the background fit function were left floating to take its flexibility given different signal hypotheses into account.
This constitutes a more rigorous consideration of background uncertainties than commonly employed in previous searches, including Ref.~\cite{EXOT-2016-20}.
In comparison to the previous approach, this results in more conservative exclusion limits that are better comparable to other recent dijet searches with similar background fits \cite{EXOT-2019-01,CMS-EXO-16-056,CMS-EXO-19-004}.
Pseudodata injected with signals of various amplitudes were used to validate proper coverage of the derived limits with all systematic uncertainties considered.

\begin{figure}[h!]
\centering
\subfloat[]{\includegraphics[width=0.49\linewidth,trim={25 0 10 0}]{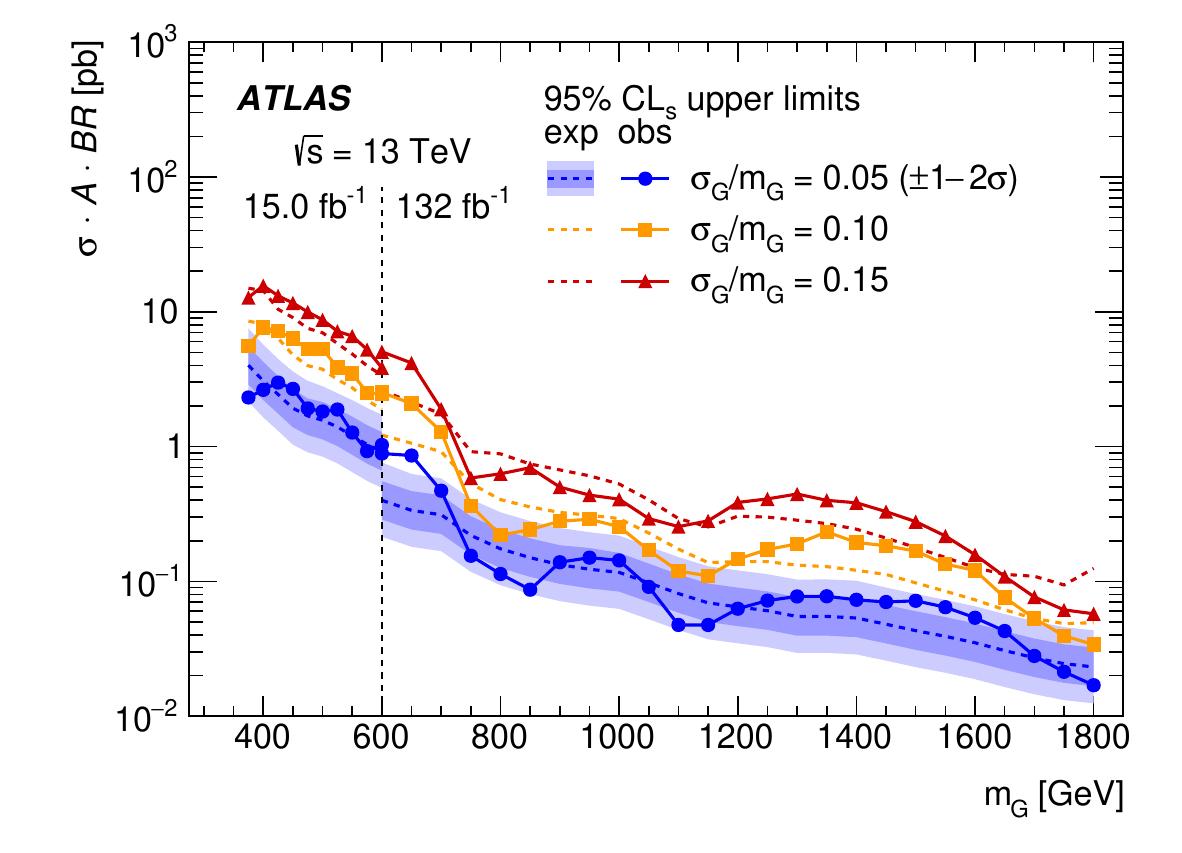}\label{fig:results_limits_gauss}}
\hfill
\subfloat[]{\includegraphics[width=0.49\linewidth,trim={25 0 10 0}]{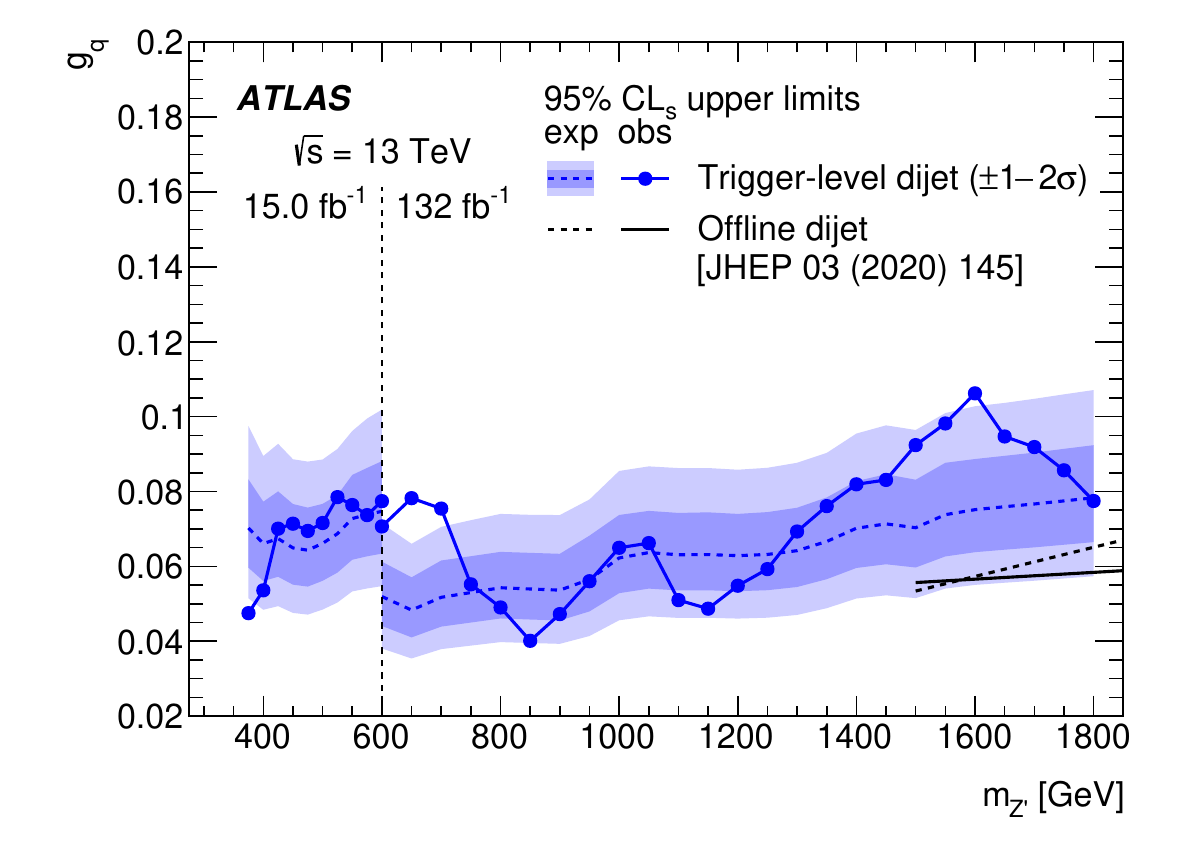}\label{fig:results_limits_zprime}}
\caption{Observed \(95\%\) \(\text{CL}_\text{S}\) upper limits on \protect\subref{fig:results_limits_gauss} the production cross-section times acceptance times branching ratio to jets, $\sigma \cdot A \cdot \text{BR}$, of  Gaussian-shaped signals of \(5\%\) (filled circles), \(10\%\) (squares) and \(15\%\) (triangles) width relative to their peak mass, $m_G$, and \protect\subref{fig:results_limits_zprime} the value of the coupling to quarks, $g_q$, of the \Zprime model as a function of the resonance mass $\MZprime$ (filled circles).
Also shown are the corresponding expected upper limits predicted for the case the \mjj distribution is observed to be identical to the background prediction in each bin (dashed lines).
For the narrowest Gaussian-shaped (5\%) and \Zprime signals, the shaded bands indicate the $1\sigma$ (darker band) and $2\sigma$ (lighter band) envelopes of outcomes expected for Poisson fluctuations around the background expectation.
The uncertainty bands for wider Gaussian-shaped signals are of similar relative size.
Overlaid on panel \protect\subref{fig:results_limits_zprime} are the observed (darker solid line) and expected (darker dashed line) upper limits from the ATLAS Run-2 search using the main stream and offline jets~\cite{EXOT-2019-03}, which benefits from tracking information and data in the tile gap region of the calorimeter.
Limits are derived independently from the J50 and J100 samples for $\mjj$ below and above $\qty{600}{\GeV}$, respectively, denoted by the vertical dashed line, as discussed in the text.
}
\label{fig:limits}
\end{figure}

\cref{fig:limits} shows the constraints on the two types of signals.
The observed exclusion limits are in good agreement with the background-only expectation.
The largest localised excess is observed for a \Zprime signal with a mass of $\MZprime=\qty{650}{\giga\electronvolt}$, corresponding to a $3.4\sigma$ local significance.
This translates to a global significance of \(2.2\sigma\) as determined using the expected number of ``downcrossings'' in pseudodata~\cite{Gross2010}.
Fluctuations in the expected limits arise from the mass-dependent spurious signal systematic uncertainty.
The model-dependent constraints in \cref{fig:results_limits_zprime} are also overlaid with results from the ATLAS Run~2 dijet resonance search using offline jets~\cite{EXOT-2019-03}, in the overlapping \Zprime mass range above \qty{1500}{\GeV}.
The limits from the offline search are stronger as it uses tracking information in the jet calibration and has an increased acceptance from including jets in the entire $|\eta|<2.4$ range.

The limits on Gaussian-shaped resonances are intended to be used to constrain signal models other than the \Zprime when PDF and non-perturbative effects can be safely truncated or neglected and, after applying the resonance selection, the reconstructed \mjj distribution predicted by the model approaches a Gaussian distribution.
For example, for sufficiently narrow resonances, the intrinsic line-shape can be neglected because the experimental \mjj resolution and parton shower tail determines the observed line-shape of the signal.
Signals with an intrinsic width much smaller than 5\% should be compared to the limit for width equal to 5\% of its mass, corresponding to 1--2 times the experimental resolution over the \mjj range considered.
Signals with larger widths should be compared with the limit that corresponds most closely to the width of the peak predicted by the model.
More instructions can be found in Appendix~A of Ref.~\cite{EXOT-2013-11}.

%
%
%

% End of text imported from the .//08limits.tex input file

\FloatBarrier

% The next lines are included from the .//09conclusions.tex input file
\section{Conclusion}
\label{sec:conclusion}

This paper describes a search for sub-\unit{\TeV} dijet resonances in \maxIntLumi of proton–proton collisions with a centre-of-mass energy of $\sqrt{s} = \qty{13}{\TeV}$ recorded by the ATLAS detector at the Large Hadron Collider.
The search was performed using the TLA technique on \onlineJetsAdjective jets to obtain up to $75$ times greater numbers of events than available to conventional searches for dijet resonances in the targeted mass range.
This search exploits an end-of-fill triggering scheme and the \LevelOneTopo topological trigger processor to extend the search region further into the electroweak mass region, from \qty{600}{\GeV} to \qty{375}{\GeV}.
This result also improves the statistical treatment of the data-derived background estimate used by the previous result~\cite{EXOT-2016-20}, allowing the high-statistics background from each sample to be modelled with a single functional fit.

The dijet invariant mass distribution recorded with the TLA technique exhibits no significant local excesses above a data-derived estimate of the smoothly falling distribution predicted by the Standard Model.
The data were used to derive 95\% confidence-level upper limits on the couplings of a leptophobic \Zprime model with axial-vector couplings to quarks, which range from $g_q<0.04$ to $g_q<0.11$ depending on \MZprime, and on the cross-section for new processes that would produce a Gaussian-shaped contribution to the dijet mass distribution.
The latter constraints exclude Gaussian-shaped contributions if the effective cross-section times acceptance times branching ratio exceeds values ranging from approximately \qty{17}{\femto\barn} to \qty{16}{\pico\barn} depending on resonance mass and width.
These results extend and improve the limits obtained from partial Run~2 data to lower production cross-sections and coupling values, superseding the previous ATLAS result~\cite{EXOT-2016-20}.
They constitute the strongest published exclusion limits on dijet resonances in the mass range \qtyrange{375}{425}{\GeV} and \qtyrange{700}{1500}{\GeV} \cite{EXOT-2019-01,CMS-EXO-14-005,EXOT-2019-03}.
%

% End of text imported from the .//09conclusions.tex input file

%
\section*{Acknowledgements}
%

% The next lines are included from the .//acknowledgements/Acknowledgements.tex input file
%
%

%
%

We thank CERN for the very successful operation of the LHC and its injectors, as well as the support staff at
CERN and at our institutions worldwide without whom ATLAS could not be operated efficiently.

The crucial computing support from all WLCG partners is acknowledged gratefully, in particular from CERN, the ATLAS Tier-1 facilities at TRIUMF/SFU (Canada), NDGF (Denmark, Norway, Sweden), CC-IN2P3 (France), KIT/GridKA (Germany), INFN-CNAF (Italy), NL-T1 (Netherlands), PIC (Spain), RAL (UK) and BNL (USA), the Tier-2 facilities worldwide and large non-WLCG resource providers. Major contributors of computing resources are listed in Ref.~\cite{ATL-SOFT-PUB-2025-001}.

We gratefully acknowledge the support of ANPCyT, Argentina; YerPhI, Armenia; ARC, Australia; BMWFW and FWF, Austria; ANAS, Azerbaijan; CNPq and FAPESP, Brazil; NSERC, NRC and CFI, Canada; CERN; ANID, Chile; CAS, MOST and NSFC, China; Minciencias, Colombia; MEYS CR, Czech Republic; DNRF and DNSRC, Denmark; IN2P3-CNRS and CEA-DRF/IRFU, France; SRNSFG, Georgia; BMFTR, HGF and MPG, Germany; GSRI, Greece; RGC and Hong Kong SAR, China; ICHEP and Academy of Sciences and Humanities, Israel; INFN, Italy; MEXT and JSPS, Japan; CNRST, Morocco; NWO, Netherlands; RCN, Norway; MNiSW, Poland; FCT, Portugal; MNE/IFA, Romania; MSTDI, Serbia; MSSR, Slovakia; ARIS and MVZI, Slovenia; DSI/NRF, South Africa; MICIU/AEI, Spain; SRC and Wallenberg Foundation, Sweden; SERI, SNSF and Cantons of Bern and Geneva, Switzerland; NSTC, Taipei; TENMAK, T\"urkiye; STFC/UKRI, United Kingdom; DOE and NSF, United States of America.

Individual groups and members have received support from BCKDF, CANARIE, CRC and DRAC, Canada; CERN-CZ, FORTE and PRIMUS, Czech Republic; COST, ERC, ERDF, Horizon 2020, ICSC-NextGenerationEU and Marie Sk{\l}odowska-Curie Actions, European Union; Investissements d'Avenir Labex, Investissements d'Avenir Idex and ANR, France; DFG and AvH Foundation, Germany; Herakleitos, Thales and Aristeia programmes co-financed by EU-ESF and the Greek NSRF, Greece; BSF-NSF and MINERVA, Israel; NCN and NAWA, Poland; La Caixa Banking Foundation, CERCA Programme Generalitat de Catalunya and PROMETEO and GenT Programmes Generalitat Valenciana, Spain; G\"{o}ran Gustafssons Stiftelse, Sweden; The Royal Society and Leverhulme Trust, United Kingdom.

In addition, individual members wish to acknowledge support from CERN: European Organization for Nuclear Research (CERN DOCT); Chile: Agencia Nacional de Investigaci\'on y Desarrollo (FONDECYT 1230812, FONDECYT 1240864, Fondecyt 3240661); China: Chinese Ministry of Science and Technology (MOST-2023YFA1605700, MOST-2023YFA1609300), National Natural Science Foundation of China (NSFC - 12175119, NSFC 12275265); Czech Republic: Czech Science Foundation (GACR - 24-11373S), Ministry of Education Youth and Sports (ERC-CZ-LL2327, FORTE CZ.02.01.01/00/22\_008/0004632), PRIMUS Research Programme (PRIMUS/21/SCI/017); EU: H2020 European Research Council (ERC - 101002463); European Union: European Research Council (BARD No. 101116429, ERC - 948254, ERC 101089007), European Regional Development Fund (SMASH COFUND 101081355, SLO ERDF), European Union, Future Artificial Intelligence Research (FAIR-NextGenerationEU PE00000013), Italian Center for High Performance Computing, Big Data and Quantum Computing (ICSC, NextGenerationEU); France: Agence Nationale de la Recherche (ANR-21-CE31-0022, ANR-22-EDIR-0002, ANR-24-CE31-0504-01); Germany: Deutsche Forschungsgemeinschaft (DFG - 469666862, DFG - CR 312/5-2); China: Research Grants Council (GRF); Italy: Istituto Nazionale di Fisica Nucleare (ICSC, NextGenerationEU), Ministero dell'Università e della Ricerca (NextGenEU 153D23001490006 M4C2.1.1, NextGenEU I53D23000820006 M4C2.1.1, NextGenEU I53D23001490006 M4C2.1.1, SOE2024\_0000023); Japan: Japan Society for the Promotion of Science (JSPS KAKENHI JP22H01227, JSPS KAKENHI JP22H04944, JSPS KAKENHI JP22KK0227, JSPS KAKENHI JP24K23939, JSPS KAKENHI JP24KK0251, JSPS KAKENHI JP25H00650, JSPS KAKENHI JP25H01291, JSPS KAKENHI JP25K01023); Norway: Research Council of Norway (RCN-314472); Poland: Ministry of Science and Higher Education (IDUB AGH, POB8, D4 no 9722), Polish National Science Centre (NCN 2021/42/E/ST2/00350, NCN OPUS 2023/51/B/ST2/02507, NCN UMO-2019/34/E/ST2/00393, UMO-2022/47/O/ST2/00148, UMO-2023/49/B/ST2/04085, UMO-2023/51/B/ST2/00920, UMO-2024/53/N/ST2/00869); Portugal: Foundation for Science and Technology (FCT); Spain: Ministry of Science and Innovation (MCIN \& NextGenEU PCI2022-135018-2, MICIN \& FEDER PID2021-125273NB, RYC2019-028510-I, RYC2020-030254-I, RYC2021-031273-I, RYC2022-038164-I); Sweden: Carl Trygger Foundation (Carl Trygger Foundation CTS 22:2312), Swedish Research Council (Swedish Research Council 2023-04654, VR 2021-03651, VR 2022-03845, VR 2022-04683, VR 2023-03403, VR 2024-05451), Knut and Alice Wallenberg Foundation (KAW 2018.0458, KAW 2022.0358, KAW 2023.0366); Switzerland: Swiss National Science Foundation (SNSF - PCEFP2\_194658); United Kingdom: Royal Society (NIF-R1-231091); United States of America: U.S. Department of Energy (ECA DE-AC02-76SF00515), Neubauer Family Foundation.

%
%

% End of text imported from the .//acknowledgements/Acknowledgements.tex input file

% The next lines are included from the .//data_availability.tex input file
\section*{Data Availability}
The public release of data supporting the findings of this article will follow the CERN Open Data Policy \cite{CERN-OPEN-2020-013}.
The values of relevant plots and tables associated with this article are stored in HEPData \cite{hepdata.161624.v1}.

% End of text imported from the .//data_availability.tex input file

%
%
%
%
%
%
%

%
\clearpage
\printbibliography
\clearpage
% ATLAS Collaboration author list
% Reference date of EXOT-2018-39 is 2025-08-01
% Author list last updated on date 10-OCT-25
% Data extracted on 10-Oct-2025 for paper reference EXOT-2018-39
% at 5:23pm
 
\begin{flushleft}
\hypersetup{urlcolor=black}
{\Large The ATLAS Collaboration}

\bigskip

\AtlasOrcid[0000-0002-6665-4934]{G.~Aad}$^\textrm{\scriptsize 104}$,
\AtlasOrcid[0000-0001-7616-1554]{E.~Aakvaag}$^\textrm{\scriptsize 17}$,
\AtlasOrcid[0000-0002-5888-2734]{B.~Abbott}$^\textrm{\scriptsize 123}$,
\AtlasOrcid[0000-0002-0287-5869]{S.~Abdelhameed}$^\textrm{\scriptsize 119a}$,
\AtlasOrcid[0000-0002-1002-1652]{K.~Abeling}$^\textrm{\scriptsize 55}$,
\AtlasOrcid[0000-0001-5763-2760]{N.J.~Abicht}$^\textrm{\scriptsize 49}$,
\AtlasOrcid[0000-0002-8496-9294]{S.H.~Abidi}$^\textrm{\scriptsize 30}$,
\AtlasOrcid[0009-0003-6578-220X]{M.~Aboelela}$^\textrm{\scriptsize 45}$,
\AtlasOrcid[0000-0002-9987-2292]{A.~Aboulhorma}$^\textrm{\scriptsize 36e}$,
\AtlasOrcid[0000-0001-5329-6640]{H.~Abramowicz}$^\textrm{\scriptsize 157}$
% ...
\bigskip
\\

% ...
$^{17}$Department for Physics and Technology, University of Bergen, Bergen; Norway.\\
% ...
$^{30}$Physics Department, Brookhaven National Laboratory, Upton NY; United States of America.\\
% ...
$^{36}$$^{(a)}$Facult\'e des Sciences Ain Chock, Universit\'e Hassan II de Casablanca;$^{(b)}$Facult\'{e} des Sciences, Universit\'{e} Ibn-Tofail, K\'{e}nitra;$^{(c)}$Facult\'e des Sciences Semlalia, Universit\'e Cadi Ayyad, LPHEA-Marrakech;$^{(d)}$LPMR, Facult\'e des Sciences, Universit\'e Mohamed Premier, Oujda;$^{(e)}$Facult\'e des sciences, Universit\'e Mohammed V, Rabat;$^{(f)}$Institute of Applied Physics, Mohammed VI Polytechnic University, Ben Guerir; Morocco.\\
% ...
$^{45}$Physics Department, Southern Methodist University, Dallas TX; United States of America.\\
% ...
$^{49}$Fakult\"{a}t Physik , Technische Universit{\"a}t Dortmund, Dortmund; Germany.\\
% ...
$^{55}$II. Physikalisches Institut, Georg-August-Universit\"{a}t G\"ottingen, G\"ottingen; Germany.\\
% ...
$^{104}$CPPM, Aix-Marseille Universit\'e, CNRS/IN2P3, Marseille; France.\\
% ...
$^{119}$$^{(a)}$New York University Abu Dhabi, Abu Dhabi;$^{(b)}$United Arab Emirates University, Al Ain; United Arab Emirates.\\
% ...
$^{123}$Homer L. Dodge Department of Physics and Astronomy, University of Oklahoma, Norman OK; United States of America.\\
% ...
$^{157}$Raymond and Beverly Sackler School of Physics and Astronomy, Tel Aviv University, Tel Aviv; Israel.\\
% ...
$^{a}$ Also at Affiliated with an institute formerly covered by a cooperation agreement with CERN.\\
$^{b}$ Also at An-Najah National University, Nablus; Palestine.\\
$^{c}$ Also at Borough of Manhattan Community College, City University of New York, New York NY; United States of America.\\
$^{d}$ Also at Center for Interdisciplinary Research and Innovation (CIRI-AUTH), Thessaloniki; Greece.\\
$^{e}$ Also at Centre of Physics of the Universities of Minho and Porto (CF-UM-UP); Portugal.\\
$^{f}$ Also at CERN, Geneva; Switzerland.\\
% ...

\end{flushleft}

% Created with Glance <Atlas.Glance@cern.ch>

%

%
%

%

%
%
%
%

\end{document}